\documentclass[]{aa}
\usepackage[english]{babel}
\usepackage{amsmath, amssymb, amsfonts}
\usepackage[]{graphicx}
\usepackage{natbib,twoopt}

\begin{document} 

 \title{Experimental validation of coronagraphic focal-plane wavefront sensing for future segmented space telescopes}
 \titlerunning{Experimental validation of coronagraphic focal-plane wavefront sensing for future segmented space telescopes}
\authorrunning{Leboulleux et al.}
  \author{Lucie Leboulleux \inst{1}, Jean-Fran\c{c}ois Sauvage \inst{2} \inst{3}, R\'{e}mi Soummer \inst{4}, Thierry Fusco \inst{2} \inst{3}, Laurent Pueyo \inst{4}, Laurent M. Mugnier \inst{3}, Christopher Moriarty \inst{5}, Peter Petrone \inst{4}, Keira Brooks\inst{4}}
  \institute{LESIA, Observatoire de Paris, Universit\'e PSL, CNRS, Universit\'e de Paris, Sorbonne Universit\'e, 5 place Jules Janssen, 92195 Meudon, France
  \and Aix Marseille Universit\'{e}, CNRS, LAM (Laboratoire d'Astrophysique de Marseille) UMR 7326, 13388, Marseille, France
  \and DOTA, ONERA, Université Paris Saclay, F-92322 Châtillon – France
  \and Space Telescope Science Institute, 3700 San Martin Drive, Baltimore, MD 21218, USA
  \and Center for Astrophysics, Harvard \& Smithsonian
  \\
             \email{lucie.leboulleux@obspm.fr}} 
            

  \abstract
   {Direct imaging of Earth-like planets from space requires dedicated observatories, combining large segmented apertures with instruments and techniques such as coronagraphs, wavefront sensors, and wavefront control in order to reach the high contrast of $10^{10}$ that is required. The complexity of these systems would be increased by the segmentation of the primary mirror, which allows for the larger diameters necessary to image Earth-like planets but also introduces specific patterns in the image due to the pupil shape and segmentation and making high-contrast imaging more challenging. Among these defects, the phasing errors of the primary mirror are a strong limitation to the performance.}
   {In this paper, we focus on the wavefront sensing of segment phasing errors for a high-contrast system, using the COronagraphic Focal plane wave-Front Estimation for Exoplanet detection (COFFEE) technique.}
   {We implemented and tested COFFEE on the High-contrast imaging for Complex Aperture Telescopes (HiCAT) testbed, in a configuration without any coronagraph and with a classical Lyot coronagraph, to reconstruct errors applied on a 37 segment mirror. We analysed the quality and limitations of the reconstructions.}
   {We demonstrate that COFFEE is able to estimate correctly the phasing errors of a segmented telescope for piston, tip, and tilt aberrations of typically $100$nm RMS. We also identified the limitations of COFFEE for the reconstruction of low-order wavefront modes, which are highly filtered by the coronagraph. This is illustrated using two focal plane mask sizes on HiCAT. We discuss possible solutions, both in the hardware system and in the COFFEE optimizer, to mitigate these issues.}
    {}


   \maketitle
\section{Introduction}
\label{s:Intro}

The direct imaging of Earth-like planets is one very exciting goal for today's astronomy. On the one hand, it could reveal the existence of life on another planet by studying the atmosphere emission lines. On the other hand, imaging a planet $10^{10}$ times fainter than its star in visible light \citep{Traub2003b} at a distance of a fraction of an arcsecond is a technological challenge. 
To do so, it requires the development of high-contrast imaging tools: starshades \citep{Brown2010, Harness2016} or coronagraphs \citep{Lyot1932,Wang1988,Malbet1996,Sivaramakrishnan2001} are used to decrease the starlight while preserving the planet light, as well as wavefront sensors and wavefront control loops \citep{Rousset1999,Fauvarque2016,Leboulleux2017,Mazoyer2018a, Mazoyer2018,Paul2014}, are used in the aim of improving the contrast in a reduced zone of the image, where the planet is located. 

From space, different solutions have been proposed through the years that present ever-increasing performance. For instance, the James Webb Space Telescope (JWST) instrument NIRCam will propose five coronagraph options to the user, including a Lyot coronagraph aiming to achieve a contrast of a few $10^{-5}$ at an inner working angle (IWA) down to $0.40$'' at $1.82$ to $2.12$ $\mu$m \citep{Krist2010}, the Wide Field Infrared Survey Telescope’s (WFIRST) Coronagraph Instrument (CGI) aims at reaching between $10^7$ and $10^9$ contrast (in most optimistic cases) around $0.1$ arcsec from the observed stars (between $0.4$ and $1$ $\mu$m) \citep{Mandell2017}, the Hubble Space Telescope (HST) a contrast around $10^{-3}$ \citep{Perrin2018}, and the Large UV/Optical/Infrared (LUVOIR) telescope instrument ECLIPS (Extreme Coronagraph for Living Planetary Systems) is designed for a $10^{-10}$ contrast between $3.5$ and $64 \lambda/D$ \citep{LUVOIR2019}. 
A combination of these tools can be seen in the WFIRST telescope \citep{Spergel2013}, the Habitable Exoplanet (HabEx) \citep{Mennesson2016} and the LUVOIR \citep{Dalcanton2015,Pueyo2017} propositions for the decadal survey. 

Among these telescopes, the LUVOIR proposition is the only one aiming for the investigation of Earth-like planets with a segmented primary mirror. The James Webb Space Telescope (JWST) \citep{Stevenson2016} and the Extremely Large Telescopes (ELTs) \citep{Macintosh2006,Kasper2008,Davies2010,Quanz2015} also include segmented primary mirrors. Such segmentation allows larger diameters: first for manufacturing reasons, and secondly, for space-based applications, to fold the telescope for launch. With a larger diameter comes a better resolution in the science image: indeed, it has been shown that the exoplanet yield increases as a power of the primary mirror diameter, especially when using a coronagraph \citep{Stark2016}. However, it also increases the complexity of these future systems since it generates additional alignment challenges: all the segments need to be well phased to fit the equivalent monolithic mirror. 

In particular, it has been proven that phasing errors of the primary mirror drastically decrease the performance of a high-contrast imaging instrument \citep{Lightsey2014, Yaitskova2003, Leboulleux2018, Stahl2015}. Such telescopes therefore require a dedicated wavefront sensor, which enables the estimation of low amplitude errors. Hardware and algorithmic solutions have to be experimentally tested and validated on optical simulators of future instruments. To respond to that need, several testbeds are emerging over the world to experimentally test and validate high-contrast imaging concepts. They are now organized in a unique online platform called the Community of Adaptive OpTics and hIgh Contrast testbeds (CAOTIC) to simplify the navigation of the user from one to the other with a fast overview and comparison of the capabilities of each testbench. Some of these testbeds include a simulation of the segmentation to confront the additional issues brought by the primary mirror segmentation, such as the Segmented Pupil Experiment for Exoplanet Detection (SPEED) at the University of Nice \citep{Martinez2018}, and the High-contrast imager for Complex Aperture Telescopes (HiCAT) testbed at the Space Telescope Science Institute (STScI) \citep{NDiaye2013}, which can simulate a 37-segment primary mirror telescope combined with a coronagraph and two deformable mirrors for wavefront control. Results have been obtained on HiCAT in the last few years, using a monolithic pupil or a well-phased segmented pupil \citep{Soummer2018}. In particular, wavefront sensing and control techniques have been deployed, such as the speckle nulling technique \citep{Soummer2018} and a parametric phase retrieval algorithm \citep{Brady2018}.

The COronagraphic Focal plane wave-Front Estimation for Exoplanet detection (COFFEE) sensor \citep{Sauvage2012a,Paul2013, Paul2013a} is capable of estimating the wavefront aberrations with high resolution. It constitutes an extension of the phase diversity technique to coronagraphic optical trains and it can be applied to a wide range of coronagraph designs, including the classical Lyot coronagraph, the Roddier \& Roddier coronagraph, the vortex coronagraph, the four quadrant phase mask coronagraph, and the apodized pupil Lyot coronagraph (APLC). This technique is highly robust to coronagraph architecture and uses the science image as the wavefront sensor, thereby making it common-path and requiring no additional hardware. It has also been validated so far on several testbeds, such as the Tr\`{e}s Haute Dynamique (THD2) testbed at the Observatoire de Paris \citep{HerscoviciSchiller2018b} and the Marseille Imaging Testbed for HIgh Contrast (MITHIC) \citep{Herscovici2019}, and on the Spectro Polarimetric High contrast Exoplanet REsearch (SPHERE) instrument \citep{Paul2013a,Paul2014a}. However, COFFEE has so far only been demonstrated on monolithic pupils.

In this paper, we propose to implement the COFFEE algorithm on HiCAT for the particular application of reconstruction of segment phasing errors, that is, local piston, tip, and tilt on the segments. One of the advantages that this test proposes is to directly use the science camera as the wavefront sensor to phase the primary mirror.

The HiCAT testbed, used for this experiment, is described in Section \ref{s:HiCATTestbedDescription}. The conditions of the experiment are then explained in Section \ref{s:ExperimentOfWavefrontSensingOnSegmentedApertures}, before developing the results without and with the coronagraph, respectively, in Sections \ref{s:ResultsWithoutCorono} and \ref{s:ResultsWithCorono}. Finally, in Section \ref{s:Conclusions}, we present our conclusions.


\section{HiCAT testbed description}
\label{s:HiCATTestbedDescription} 

Since 2013, HiCAT has been under development at the Russel B. Makidon Laboratory at STScI to investigate high-contrast technologies for future space missions \citep{NDiaye2013,NDiaye2014a,NDiaye2015a,Leboulleux2016,Leboulleux2017,Soummer2018,Brady2018}. 

It is designed to develop, test, and validate solutions for future space telescopes with complex apertures dedicated to high-contrast imaging, such as WFIRST or LUVOIR. To do so, it is aimed at allowing for the implementation of multiple different telescope configurations. The simulation of telescope pupil's complexity can include the central obstruction, the spiders, the segment gaps, and the phasing errors. These components can be easily replaced with new components, depending on the configuration tested.

To perform high-contrast imaging, HiCAT includes a starlight and diffraction suppression system (Lyot coronagraphs or APLC) and wavefront sensing and wavefront control tools (deformable mirrors and phase retrieval camera). Some of these elements can also be easily replaced using external metrology as required.

\subsection{Optical and opto-mechanical design}

\begin{figure*}
   \begin{center}
   \begin{tabular}{c}
   \includegraphics[width=10cm]{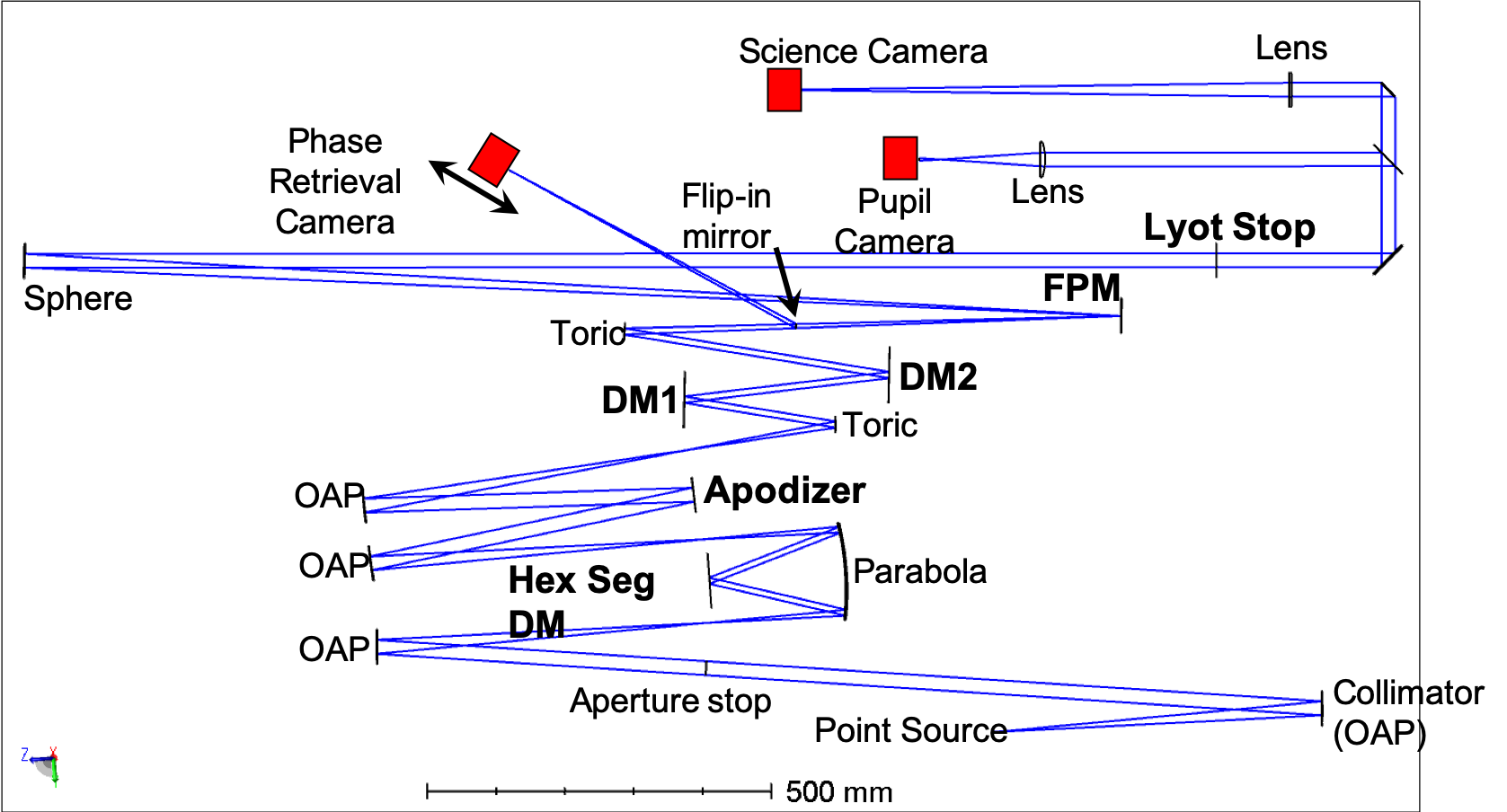}
   \end{tabular}
   \end{center}
   \caption[HiCAT] 
   { \label{fig:HiCAT} 
Optical and mechanical design of the HiCAT testbed. The point source and the collimator simumlate a star located at infinity. The aperture stop corresponds to a pupil mask and can be circular, contain spiders or be shaped such as segmented telescopes. It is conjugated is the Iris AO, here called Hexagonal Segmented DM and together, they can simulate a segmented primary mirror at its full complexity. The apodizer (pupil plane), the FPM (focal plane), and the Lyot stop form a full APLC, which can easily be downgraded to a classical Lyot corongaraph when the apodizer is replaced with a flat mirror or even be removed if no coronagraph is needed. DM1 (pupil plane) and DM2 (out of pupil plane) are used for wavefront control to correct for wavefront aberrations or apply a diversity phase. At the end of the optical beam, two cameras can be found: a pupil plane camera and the science camera, which is also used as the post-coronagraph wavefront sensor when focal plane wavefront sensing techniques are tested. The phase retrieval camera offers an alternative as a pre-coronagraph detector.}
   \end{figure*}

The final layout is presented in Figure \ref{fig:HiCAT}: the testbed includes a star simulator, an observing telescope, a coronagraph, a wavefront control stage, and a high-contrast imaging system. The pupil mask, the Iris-AO segmented mirror, the apodizer, the first deformable mirror, and the Lyot stop are located in pupil planes, while the focal plane mask (FPM) and the science camera are in focal planes. It is a purely reflective testbed, except for the final lens ahead of the cameras (after the Lyot Stop). It is also designed to minimize the impact of its optical components on the final image and contrast by reducing the sources of amplitude-induced errors from the propagation of out-of-pupil surfaces (Talbot effects) \citep{NDiaye2013}. With this set up, the majority of the amplitude errors comes from the discontinuities in the pupil which are corrected using apodization and wavefront control.

The star is simulated using a fiber source collimated with an off-axis parabolic mirror (OAP). The source can be either a monochromatic light ($\lambda = 638$ nm) or a broadband supercontinuum light source that can be combined with either a filter wheel or a monochromator ($600$ to $680$ nm).

\begin{figure}
   \begin{center}
   \begin{tabular}{c}
   \includegraphics[height=5cm]{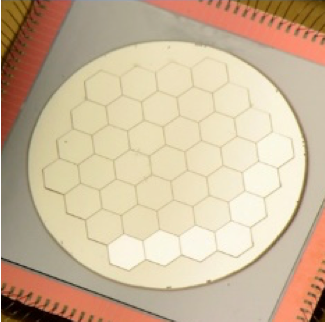}
   \end{tabular}
   \end{center}
   \caption[IrisAO] 
   { \label{fig:IrisAO} 
Picture of the Iris-AO MEMs segmented mirror set up on the HiCAT testbed and used for the experiment. It is composed of 37 hexagonal segments that can each be controlled in piston, tip, and tilt. The gaps between the segments measure between $10$ and $12 \mu$m (Iris-AO spreadsheet) and the full segmented mirror has an inscribed circle diameter of $7$ mm \citep{Soummer2018}.}
   \end{figure}

The telescope is simulated using an entrance pupil mask, conjugated with a mirror that can be either monolithic or segmented. In the latter case, we use a 37-segment Iris-AO MEMs DM with hexagonal segments that can be individually controlled in piston, tip, and tilt (see Figure \ref{fig:IrisAO}).

The coronagraph is an APLC, where the apodizer can be replaced by a flat mirror to obtain a classical Lyot coronagraph. The apodizer is easily replaceable and a couple of apodizers have already been tested: a WFIRST-like configuration (monolithic mirror, WFIRST central obstruction and spiders, and apodization optimized for this pupil) and a LUVOIR-like configuration (36 hexagonal segments, hexagonal central obstruction, spiders, apodization). The FPM is made of a reflective golden surface with a circular central hole, two diameters of holes being possible. The Lyot stop is a transmissive circular mask. Both the FPM and the Lyot stop are motorized. The wavefront control can be done with either one or two DMs: DM1 is set in a pupil plane while DM2 is out of pupil plane, to correct for both amplitude and phase aberrations. The DMs are Boston Micromachines deformable mirrors (kilo-DM), each of them with 952 actuators in a 9.9 mm (34 actuators) diameter disk. Both DMs are currently installed, but only DM2 is used since the calibration of DM1 needs to be redone. For the validation of COFFEE, we use the science camera (ASI1600mm from ZWO) with $4656 \times 3520$ pixels, each of them having a size of $3.8$ $\mu$m.

Apart from DM1 and DM2 best flats, the internal wavefront error of HiCAT, measured with a Fizeau interferometer, is around 13 nm root-mean-square (RMS). After adding and controlling the DMs, the wavefront error goes down around 2 nm \citep{Brady2018,Soummer2018}.

\subsection{Environment constraints}

For high-contrast imaging, performance is degraded by wavefront perturbations resulting from dust, turbulence, or vibrations. This is why HiCAT is mounted on a floating table, itself on a stage independent from the rest of the building to mitigate vibration effects. A box covers all the testbed to protect it from dust, particles, air flows, and parasitic light. The box is located in a class 1000 clean room (ISO 6) with temperate control better than $1^\circ$C, and humidity that is maintained around $30\%$. Additional constraints are introduced by the presence of the DMs, since they need to remain below $30\%$ of humidity without any air flow. Inside the box, temperature and humidity sensors have been installed and a complementary dry air supply reduces the humidity to well below $30\%$.

\subsection{Latest results on HiCAT}

HiCAT has the capacity to simulate both monolithic and segmented telescopes, which can be combined with a Lyot coronagraph or even a full APLC. A pair-wise estimation (wavefront sensing) and stroke minimization (wavefront control) combined procedure has been implemented and can be run in any given configuration, which provided the first results of coronagraph/wavefront control combination \citep{Soummer2018}. In monochromatic light, with an APLC coronagraph and a segmented entrance pupil, contrasts of $4 \times 10^{-7}$ and $10^{-7}$ have been reached respectively for a symmetric dark hole and a half dark hole. With a classical Lyot coronagraph replacing the APLC, a contrast of $5 \times 10^{-8}$ has been obtained, still in monochromatic light \citep{Soummer2019}. For this study, we utilize a segmented aperture and a classical Lyot coronagraph.


\section{Objective and conditions of the experiment}
\label{s:ExperimentOfWavefrontSensingOnSegmentedApertures}

\subsection{Objective of the experiment}

Segmentation of the telescope primary mirrors generates various issues, including an increasing complexity of the alignment process. In addition to the global alignment of the primary mirror, each segment has at least three degrees of freedom (piston, tip, tilt) that need to be aligned to achieve the target performance. Numerous wavefront sensors have been developed to reconstruct the wavefront aberrations and can be applied to segment phasing, such as the Zernike sensor for Extremely Low-level Differential Aberrations (ZELDA) \citep{NDiaye2013a}, the Self-Coherent Camera (SCC) sensor \citep{Baudoz2006, Mazoyer2013, Martinez2016} or the Phase Diversity \citep{Gonsalves1982, Mugnier2006}, or the ZErnike Unit for Segment phasing (ZEUS) \citep{Dohlen2006}. In high-contrast imaging, focal plane sensors are preferable, since they enable the retrieval of all aberrations from the entrance pupil to the science camera, while pupil plane wavefront sensors' reconstructions suffer from non-common path aberrations.

The COFFEE sensor only requires focal plane images, which means the science detector is used as the sensor and no additional device is needed. Another advantage of COFFEE is that it can operate in the presence of a coronagraph, which makes it a strong candidate for high-contrast imaging instruments. It has already been validated on monolithic apertures on SPHERE \citep{Paul2013, Paul2014a}, on the MITHIC testbed, and more recently on the THD2 testbed \citep{HerscoviciSchiller2018}. However, COFFEE has never been validated on bench, nor in simulation, on a segmented aperture. This is the motivation behind the experiment introduced in this paper: using COFFEE on segmented apertures to reconstruct phasing errors in the presence of a coronagraph. 

\subsection{Conditions and tools}

The optical configuration of the HiCAT testbed for this experiment and, in particular, the numerical parameters useful for the model implemented in the COFFEE algorithm, is explained in Section~\ref{sec:OpticalConditions}, and the main software steps of the experiment, from the images taken by the camera to the processed data, are detailed in Sections~\ref{sec:ImagesAndPreProcessing}, \ref{sec:PhaseReconstructionWithCOFFEE}, and \ref{sec:PostProcessing}.

\subsubsection{Optical conditions}
\label{sec:OpticalConditions}

The chosen source is a monochromatic light with a wavelength of $\lambda = 638$ nm. The entrance pupil is a circular mask of diameter $18$ mm, and the Lyot stop a circular pupil of diameter $15$ mm. Since the pupil masks hides the outer ring of the Iris-AO, the number of fully visible segments reduces from 37 to 19. The system can be used in coronagraphic imaging or in direct imaging (defined here as the imaging without FPM) by moving the FPM one centimeter away from the optical axis. One centimeter is enough to remove entirely the FPM from the optical path.

The sampling factor is an important parameter for COFFEE. It enables us to quantify the pixel scale, which means the Full Width at Half Maximum (FWHM) of the PSF with respect to the number of pixels. To estimate it, we compute the cut-off frequency $D/\lambda$ of the Modulation Transfer Function (MTF) of the system (modulus of the Fourier Transform of the PSF without FPM). The dimension of the PSF image (in pixels) divided by the cutoff frequency of the MTF (also in pixels) provides the sampling $[\lambda/D]_{pix}$ in pixels on the camera:
\begin{equation}
    [\lambda/D]_{pix} = \frac{\text{dim}(PSF)}{f_c}.
\end{equation}
As a comparison, the sampling is equal to $2$ pixels per $\lambda/D$ for Shannon-Nyquist sampled images. On HiCAT, the sampling on the camera is $13.3 \pm 0.1$ pixels per $\lambda/D$ when the Lyot stop is set up, while $11.4 \pm 0.1$ pixels per $\lambda/D$ when it is out. The ratio of these two values provides an estimation of the Lyot ratio: $85.6\% \pm 1.1\%$, more precise than the value deduced from the ratio between the Lyot stop and entrance pupil diameters.

\begin{figure}
   \begin{center}
   \begin{tabular}{cc}
   \includegraphics[height=5cm]{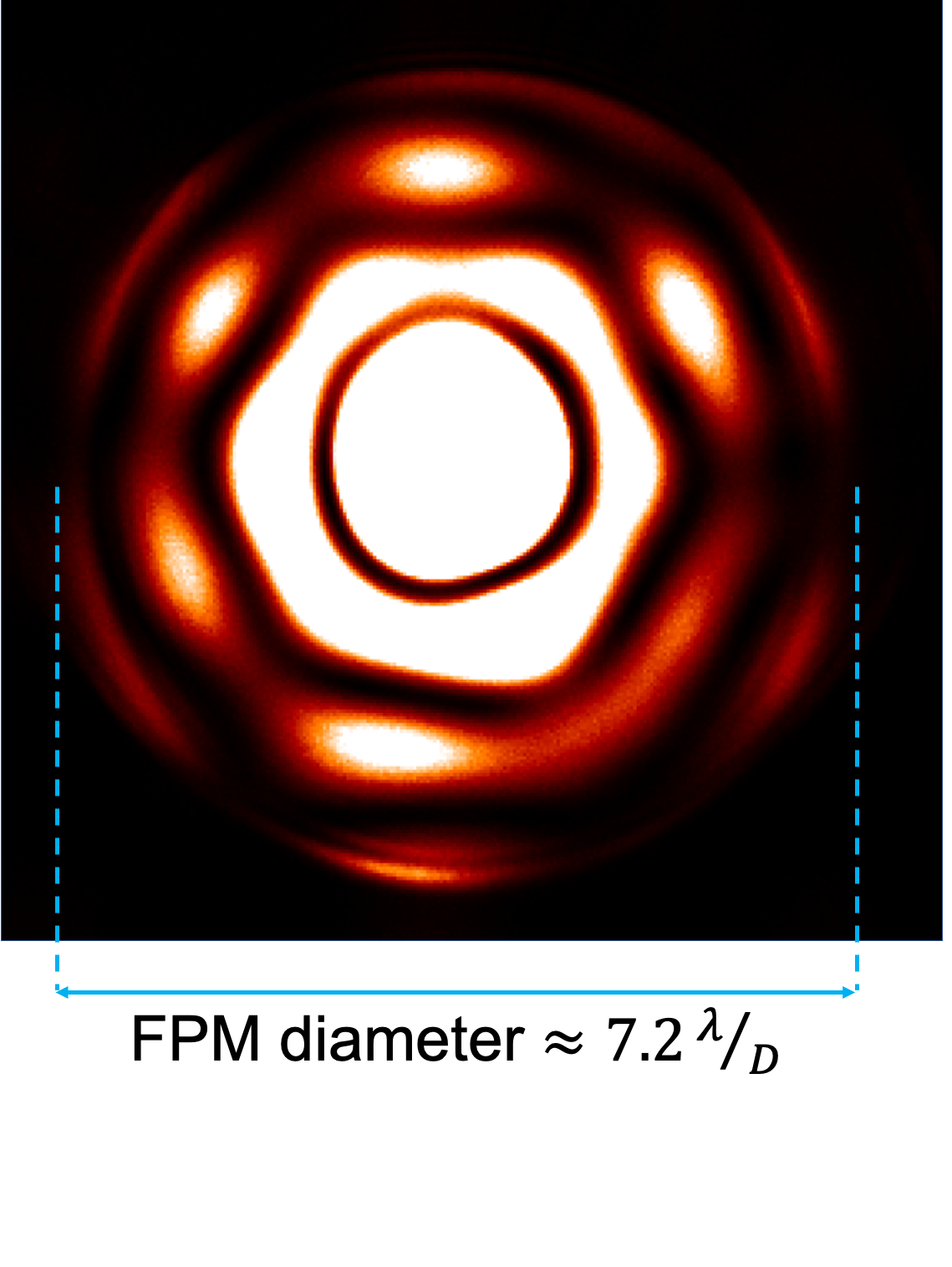} & \includegraphics[height=5cm]{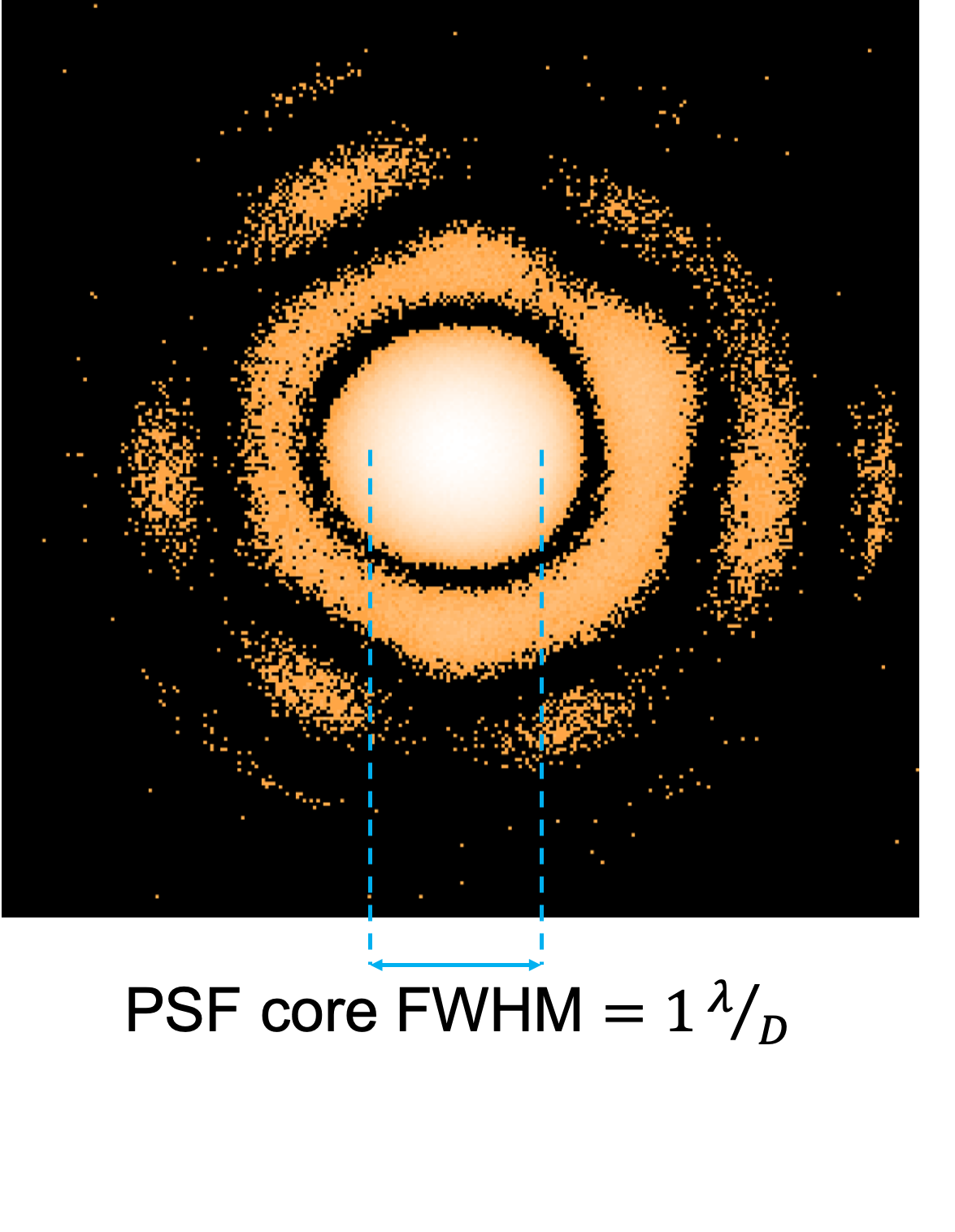} \\
   a & b
   \end{tabular}
   \end{center}
   \caption[COFFEE_FPM] 
   { \label{fig:COFFEE_FPM} 
FPM size calibration: Images of the PSF at the FPM plane, cropped by the mask. (a) Saturated image (linear scale) and (b) non-saturated image (logarithmic scale). In this case, the FPM diameter is estimated at $7.2 \pm 0.9 \lambda/D$.}
   \end{figure}

Since COFFEE relies on a fine model of the imaging system, the FPM size in $\lambda/D$ has also to be known with precision. On HiCAT, the FPM is a hole in a flat mirror. To compute a precise estimation of its diameter, we install a camera in a focal plane behind it, with the detector conjugated to the FPM. We can then observe a PSF cropped after around two rings by the mask. Two different light exposures are used: by increasing the source intensity until saturating the inside of the hole (Figure \ref{fig:COFFEE_FPM}-a), we can deduce a precise estimation in pixels of the hole diameter; and by adjusting the source intensity to obtain a non-saturated PSF (Figure \ref{fig:COFFEE_FPM}-b), we can measure its FWHM, which provides an estimation of the resolution element $\lambda/D$ in pixels. By dividing our precise estimate of the hole diameter in pixels by the FWHM estimate of the resolution element, we obtain a diameter of $7.2 \pm 0.9 \lambda/D$ for the FPM:
\begin{equation}
    [D_{FPM}]_{\lambda/D} = \frac{[D_{FPM}]_{pix}}{[\lambda/D]_{pix}} \quad \text{where} \quad [\lambda/D]_{pix} = [FWHM_{PSF}]_{pix}.
\end{equation}
This FPM extent also provides the cutoff frequency in coronagraphic mode for the rest of the experiment, which will be discussed later in this paper.

\subsubsection{Image acquisition and pre-processing}
\label{sec:ImagesAndPreProcessing}

\begin{figure}
   \begin{center}
   \begin{tabular}{c}
   \includegraphics[width=8.5cm]{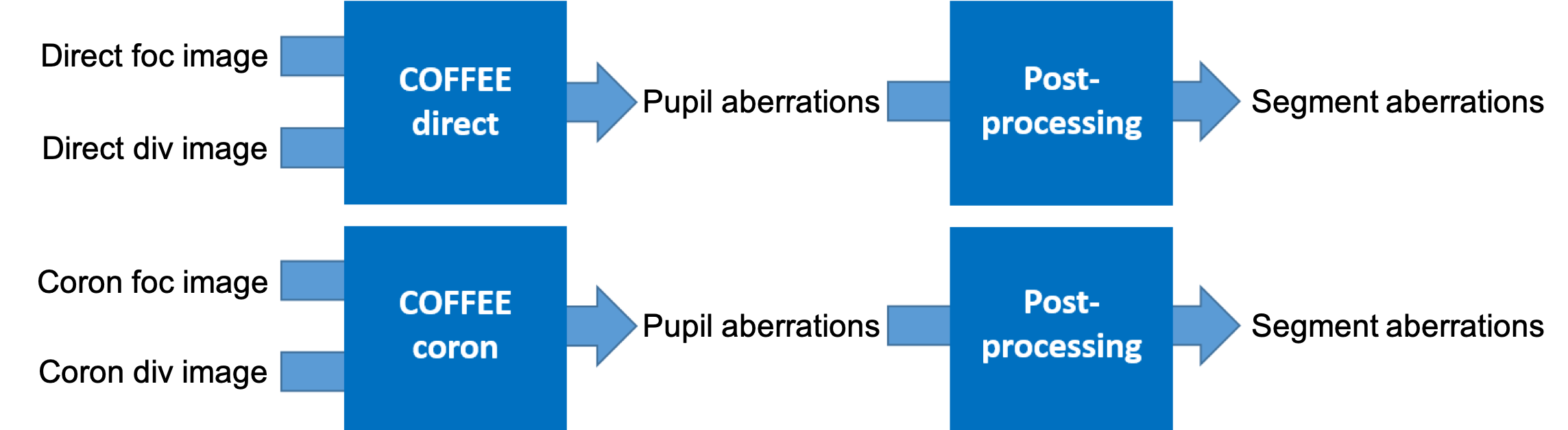}
   \end{tabular}
   \end{center}
   \caption[COFFEE_scheme] 
   { \label{fig:COFFEE_scheme} 
Structure of the data analysis process. "coron" stands for "coronagraphic", "bd" for "background", "foc" for "on focus" (the DM is flat), and "div" for "diversity". The COFFEE reconstruction is done independently but with the same method in both configurations (direct and coronagraphic) for comparisons later on. The segment pistons extracted in both cases will also be compared to the commands sent to the Iris-AO segmented mirror.}
   \end{figure}

COFFEE requires two images: a focus image and a diversity image (Figure \ref{fig:COFFEE_scheme}). In our case, the diversity image is obtained with applying a $300$ nm Peak-to-Valley (PV) focus on the pupil plane DM (the focus was calibrated in front of a Fizeau interferometer with a precision of a few nanometers), while the DM remains flat for the focus image.

Figure \ref{fig:COFFEE_scheme} describes the successive pre-processing steps of this experiment. Pre-processing is done to increase the quality of the input images of COFFEE and reduce the noise. Each input image corresponds to the average of 400 images with the average of 400 background images subtracted. The background images are taken when a motorized light trap blocks the beam a few centimeters ahead of the FPM. All images have $1424\times1424$ pixels with a sampling of either $11.4 \pm 0.1$ or $13.3 \pm 0.1$ pixels per $\lambda/D$ depending if the Lyot stop is in or out of the optical path (see Section~\ref{sec:OpticalConditions}). These steps are the same in both direct and coronagraphic modes.

All these configurations (different DM phase application, FPM position control, light suppression component control, camera use, and image pre-processing) require a safe hardware control. On HiCAT, it has been developed in object-oriented Python as a generic package: each hardware is a child class of a generic "instrument" parent class, and each action is a function. Each electronic on the testbed can be controlled remotely and safely, enabling to switch from a mode to another or to put on or off the light trap without any interruption. The entire sequence from hardware control and image acquisition to image pre-processing belongs to this unified and generic software infrastructure \citep{Moriarty2018}.

\subsubsection{Phase diversity with COFFEE}
\label{sec:PhaseReconstructionWithCOFFEE}

The COFFEE algorithm is an extension of the phase diversity technique to coronagraphic images developed by \cite{Sauvage2012a} and \cite{Paul2013}. It is a generic code that works for different coronagraphs such as the APLC, the classical Lyot coronagraph, the Roddier \& Roddier coronagraph, the four quadrant phase mask, or even no coronagraph. The model of image formation used in COFFEE includes a coronagraph with all its components : apodizer, FPM, and Lyot stop.

Two coronagraphic images, $I_f$ and $I_d$, are needed. We obtain $I_d$ after applying a known diversity phase $\phi_d$ in the entrance pupil plane, using the deformable mirror. COFFEE enables us to estimate two unknown phase aberrations: $\phi_{up}$ corresponds to the upstream phase aberrations, that is, before the coronagraph (typically in the apodizer plane) and $\phi_{do}$ to the downstream aberrations, that is, after the coronagraph (typically in the Lyot stop plane). We call $E_0$ the amplitude of the electric field. The coronagraph function $\mathcal{C}\{E_0e^{i\phi_{up}},e^{i\phi_{do}}\}$ expresses the propagation from each plane to the following one (with a Fourier Transform) and the multiplications by the corresponding masks. We also consider residual continuous backgrounds in the images, $\beta_f$ and $\beta_d$ due to the coronagraph, and the noises (detector and photon noises) of $I_f$ and $I_d$ respectively, $n_f$ and $n_d$, such as:
\begin{equation} 
    I_f = \left \Vert \mathcal{C}\{E_0e^{i\phi_{up}},e^{i\phi_{do}}\}\right \Vert ^2 + n_f + \beta_f,
\end{equation}
\begin{equation} 
    I_d = \left \Vert \mathcal{C}\{E_0e^{i\phi_{up}+i\phi_d},e^{i\phi_{do}}\}\right \Vert ^2 + n_d + \beta_d.
\end{equation}
The objective of the COFFEE algorithm is to maximize $p(E_0,\beta_f,\beta_d,\phi_{up},\phi_{do}|I_f,I_d)$, the probability that we may observe the parameters $E_0$, $\beta_f$, $\beta_d$, $\phi_{up}$, and $\phi_{do}$, knowing the images $I_f$ and $I_d$. It is equivalent to minimizing the criterion $J_{MAP}=-\ln(p(E_0,\beta_f,\beta_d,\phi_{up},\phi_{do}|I_f,I_d))$. This criterion has been made analytically explicit in \cite{Sauvage2012a} and \cite{Paul2013}. Then, it typically calls for a numerical minimization based on
the variable metric with limited memory and bounds (VMLM-B) method [\cite{Paul2014}]. The minimization stops when the difference between two successive values of the criterion is below a certain threshold, fixed by the user, and COFFEE provides the best estimates of $E_0$, $\beta_f$, $\beta_d$, $\phi_{up}$, and $\phi_{do}$ according to this minimization.

COFFEE is particularly well suited for our study, since the goal is to measure the wavefront aberrations in a coronagraphic system, with no differential aberration and as small modifications of the system as possible. Secondly, it provides the upstream phase $\phi_{up}$, which corresponds, in our experiment, to the surface of the segmented mirror. Finally, it reconstructs the aberrations on a pixel-wise basis (in opposition for instance with a Zernike-wise basis), which means that the phase function itself is estimated. It is therefore a good candidate for segment phasing, which requires a high spatial resolution.

\subsubsection{Post-processing}
\label{sec:PostProcessing}

\begin{figure}
   \begin{center}
   \begin{tabular}{c}
   \includegraphics[width=8.5cm]{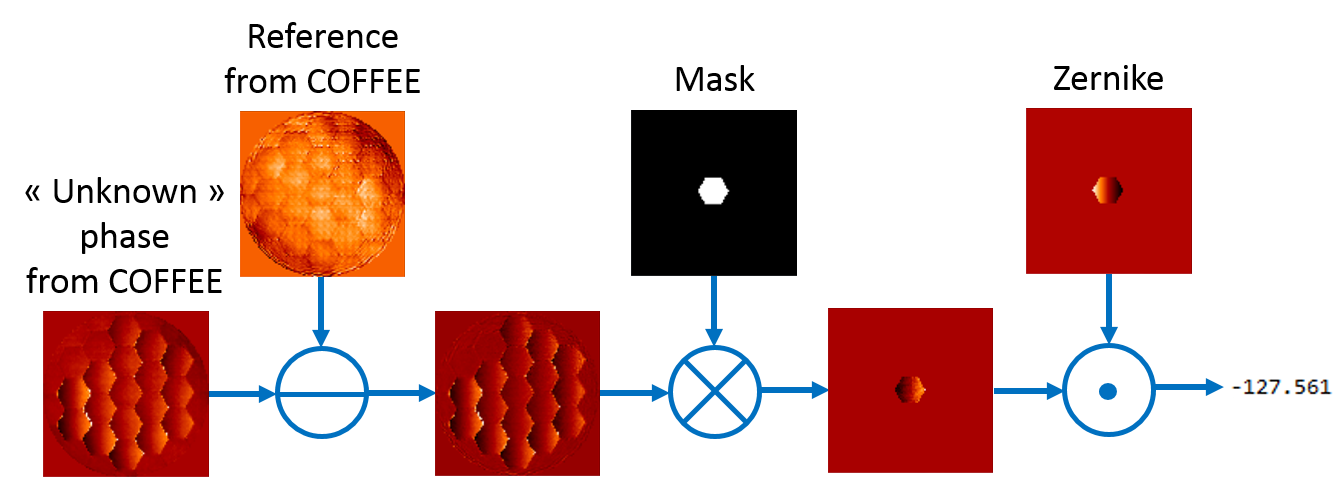}
   \end{tabular}
   \end{center}
   \caption[PostProcessing] 
   { \label{fig:PostProcessing} 
Post-processing to deduce the piston, tip, and tilt values on the different segments from the phases reconstructed by COFFEE. First, we subtract the reference phase (obtained with COFFEE when the Iris-AO is at its best flat) to obtain the residual piston/tip/tilt to be corrected. The residual phase is multiplied with a hexagonal mask to isolate the segment of interest. We apply a dot product to this segment phase with a hexagonal calibrated Zernike polynomial to compute the aberration amplitude, that can be compared to the command that has been sent to the Iris-AO. The output value, here $-127.561$ is in nanometers.}
   \end{figure}

To obtain the phasing errors per segment the upstream phase aberration map provided by COFFEE is projected on local segment-level piston, tip, and tilt modes. In our case, we want to verify if the output piston and tip-tilt values correspond to the command sent to the Iris-AO mirror. To do so, we follow the protocol described in Figure \ref{fig:PostProcessing}: from the map deduced by COFFEE, we subtract a reference phase, reconstructed by COFFEE but with the Iris-AO at its best flat. This enables the removal of global aberrations and reconstruction artifacts such as the global tip-tilt, that do not interest us in this experiment. Then a hexagonal mask is put on each segment to isolate it and get the aberration of the segment only. This isolated aberration is projected on tip mode normalized at 1nm RMS with a dot product. To get the piston value, this last step is replaced by computing the average of the phase inside the mask. The output values then correspond to the Zernike coefficients of the segment-level phase.

\section{Reconstruction of phasing errors without coronagraph}
\label{s:ResultsWithoutCorono}

In this section, we focus on determining the linearity of COFFEE without the FPM. COFFEE is then equivalent to a phase diversity algorithm \citep{Gonsalves1982}, that is, when the coronagraph model is a single Fourier Transform. The objective is to set reference results with which we will compare the results obtained in presence of the coronagraph.

\subsection{Examples of reconstructed phases}
\label{s:ExamplesOfReconstructedPhases}

\begin{figure}
   \begin{center}
   \includegraphics[width=8.5cm]{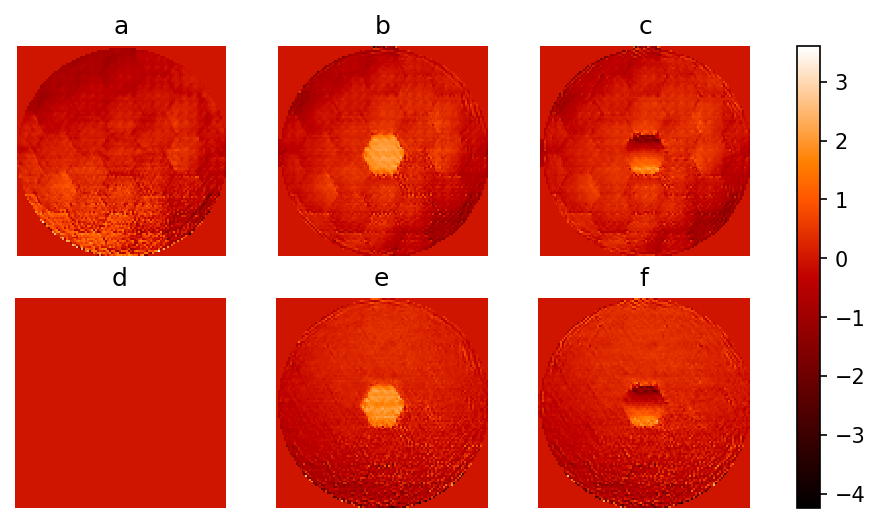}
   \end{center}
   \caption[direct_phase_COFFEE] 
   { \label{fig:direct_phase_COFFEE} 
Phases reconstructed with COFFEE in direct (non-coronagraphic) configuration, in radians: (a) when the Iris-AO is flattened (this map is also used as our reference phase), (b) when a $188$ nm piston is applied on the central segment (open loop), (c) when a $0.2$ mrad tip wavefront is applied on the central segment. (d), (e), and (f) corresponds to respectively (a), (b), and (c) after subtraction of the reference phase (phase estimated by COFFEE when the Iris-AO mirror is flattened, that is, (a)). The aberrations are well reconstructed by COFFEE, both for low order aberrations which are similar in all reconstructions and in high frequency aberrations, which correspond to the borders of the segments.}
   \end{figure}

Figure \ref{fig:direct_phase_COFFEE} introduces some phases reconstructed with COFFEE in direct configuration (no coronagraph), when known aberrations are applied on the segmented mirror. On the first column (a), the segmented mirror is flattened, the reconstructed phase corresponds to the so-called reference phase. On the second column (b), the Iris-AO is flat except for the central segment on which a $188$ nm piston is applied. On the third column (c), the Iris-AO is also flat, except for the central segment that has a $0.1$ mrad tip. The first line (1) contains the raw phase maps as computed by COFFEE, while the second line (2) corresponds to the raw phase maps minus the reference phase (1a).

In all cases, the subtraction of the reference phase removes most of the global aberrations and reduces the segmentation pattern on flattened areas, that is, the edge effects.

\subsection{Linearity of phase diversity to segment aberrations}
\label{s:PrecisionAnalysisLimits}

\begin{figure*}
   \begin{center}
   \begin{tabular}{cc}
   \includegraphics[height=6.5cm]{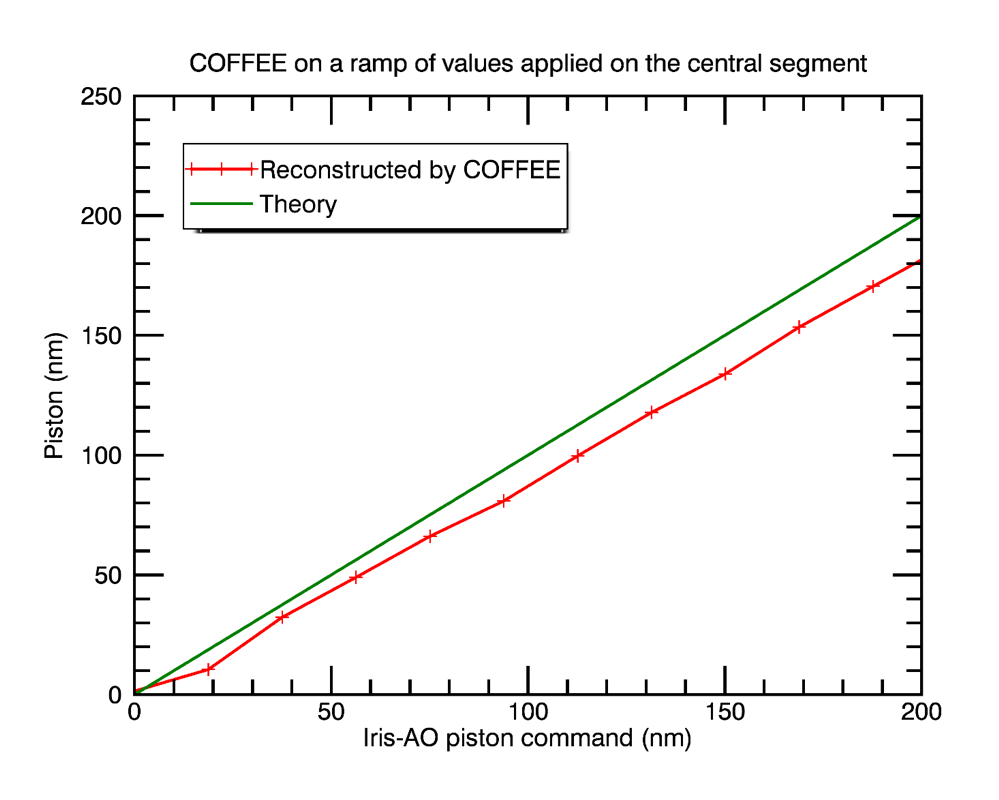} &
   \includegraphics[height=6.5cm]{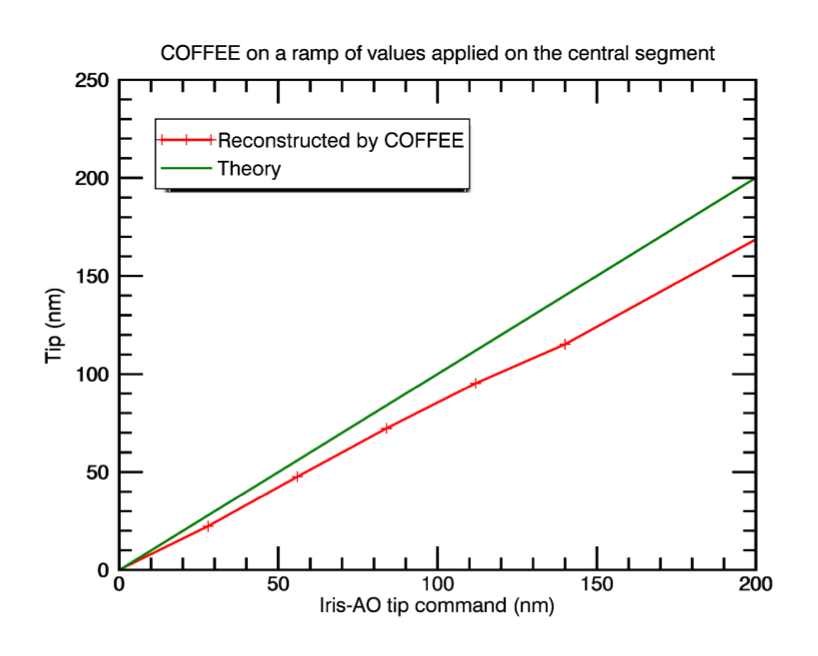} \\
   a & b
   \end{tabular}
   \end{center}
   \caption[Piston_ramp_direct] 
   { \label{fig:Piston_ramp_direct} 
(a) Linearity of the phase diversity on the estimation of a piston aberration on the central segment (range of piston values from 0 to 200 nm), in the direct case (no coronagraph). (b) Linearity of the phase diversity on the estimation of a tip aberration on the central segment (range of values from 0 to 200 nm pv). In both cases, the theory curves correspond to $y=x$. In the piston case (a) the curve can be approximated with the linear function $y = 0.932 (\pm 0.006) x -4.02 (\pm 0.97)$ and in the tip case (b) it can be approximated with the linear function $y = 0.857 (\pm 0.009) x -0.98 (\pm 1.2)$. These plots show a possible uncertainty in the calibration of Iris-AO.}
   \end{figure*}

The objective here is to study the error of reconstruction of COFFEE for different kinds of phasing aberrations: piston, tip, and tilt. To study piston aberrations, the Iris-AO remains flat, except for the central segment on which different piston values are applied: from $0$ to $200$ nm. We estimate with COFFEE the aberration phase for a set of images, and deduce the respective piston value of the central segment. This set of values is compared to the command sent to the central segment of the Iris-AO. Figure \ref{fig:Piston_ramp_direct} (a) illustrates this study. 

We process similarly for the study of tip aberrations: the Iris-AO is still flat, except for the central segment on which seven tip commands from 0 to 200 nm are successively applied. The results are indicated on Figure \ref{fig:Piston_ramp_direct} (b).

\subsection{Result analysis}
\label{s:DirectConfiguration}

For the piston case, the curve looks like the linear function $y = 0.932 (\pm 0.006) x -4.02 (\pm 0.97)$ (in nanometers). The RMS of the difference between the linear fit and the data from COFFEE is $1.97$ nm. For the tip case, the curve looks again linear: $y = 0.857 (\pm 0.009) x -0.98 (\pm 1.2)$ (in nanometers). The RMS error between all data points and this fit is $1.68$ nm.

In these two cases, we notice that even if the estimation is consistent with the command sent to the Iris-AO, the Zernike coefficients estimated by COFFEE and our post-processing algorithm are a bit underestimated. We hypothesize that the difference between the COFFEE values and the expected values increasing with the level of the aberration applied on the Iris-AO mirror, could be due to an overestimation of the Iris-AO calibration coefficients, since they are sent in open loop and the commands are not calibrated using for example an interferometer. Furthermore, the non-null y-intercepts indicate that COFFEE detects in both cases a default slight phasing error on the Iris-AO.

\section{Reconstruction of phasing errors with a coronagraph}
\label{s:ResultsWithCorono}

In this section, we focus on the data taken with a $7.2 \lambda/D$-large FPM and a $85.6\% \pm 1.1\%$ Lyot stop and estimated with COFFEE.

\subsection{Observations}
\label{s:Observations}

We proceed with the same strategy than in the previous section, except for the coronagraph model used in COFFEE and the FPM centered on the optical path. 

\begin{figure}
   \begin{center}
   \includegraphics[width=8.5cm]{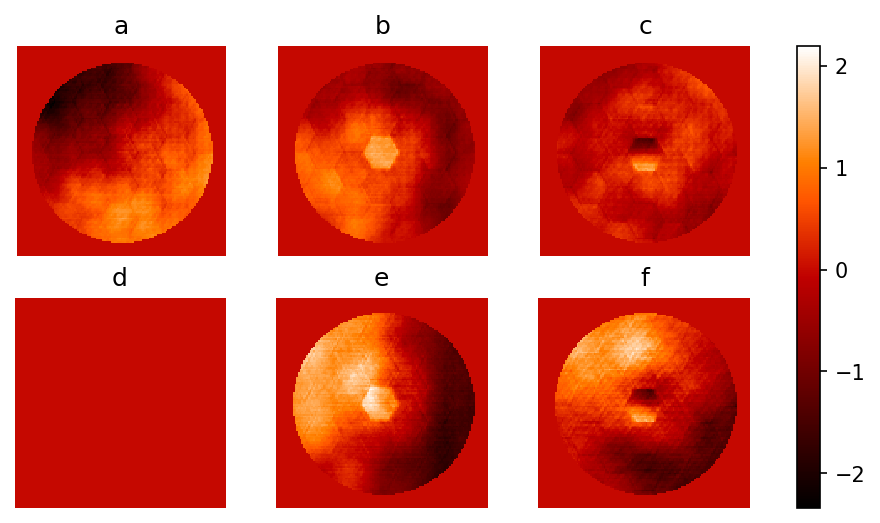}
   \end{center}
   \caption[coron_phase_COFFEE] 
   { \label{fig:coron_phase_COFFEE} 
Phases reconstructed with COFFEE in coronagraphic configuration, in radians: (a) when the Iris-AO is flattened (it is also our reference phase), (b) when a $188$ nm piston is applied on the central segment, (c) when a $0.2$ mrad tip is applied on the central segment. (d), (e), and (f) corresponds to respectively (a), (b), and (c) after subtraction of the reference phase (phase estimated by COFFEE when the Iris-AO mirror is flattened, that is, (a). Even if high-order seem quite well reconstructed, low-order aberrations in the three cases are not equally reconstructed.}
   \end{figure}

\begin{figure*}
   \begin{center}
   \begin{tabular}{cc}
   \includegraphics[height=5cm]{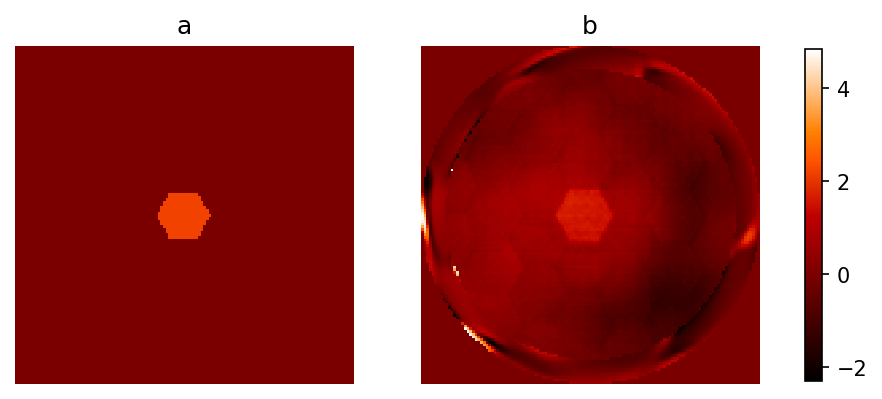}
   & \hspace{-0.7cm} \includegraphics[height=5cm]{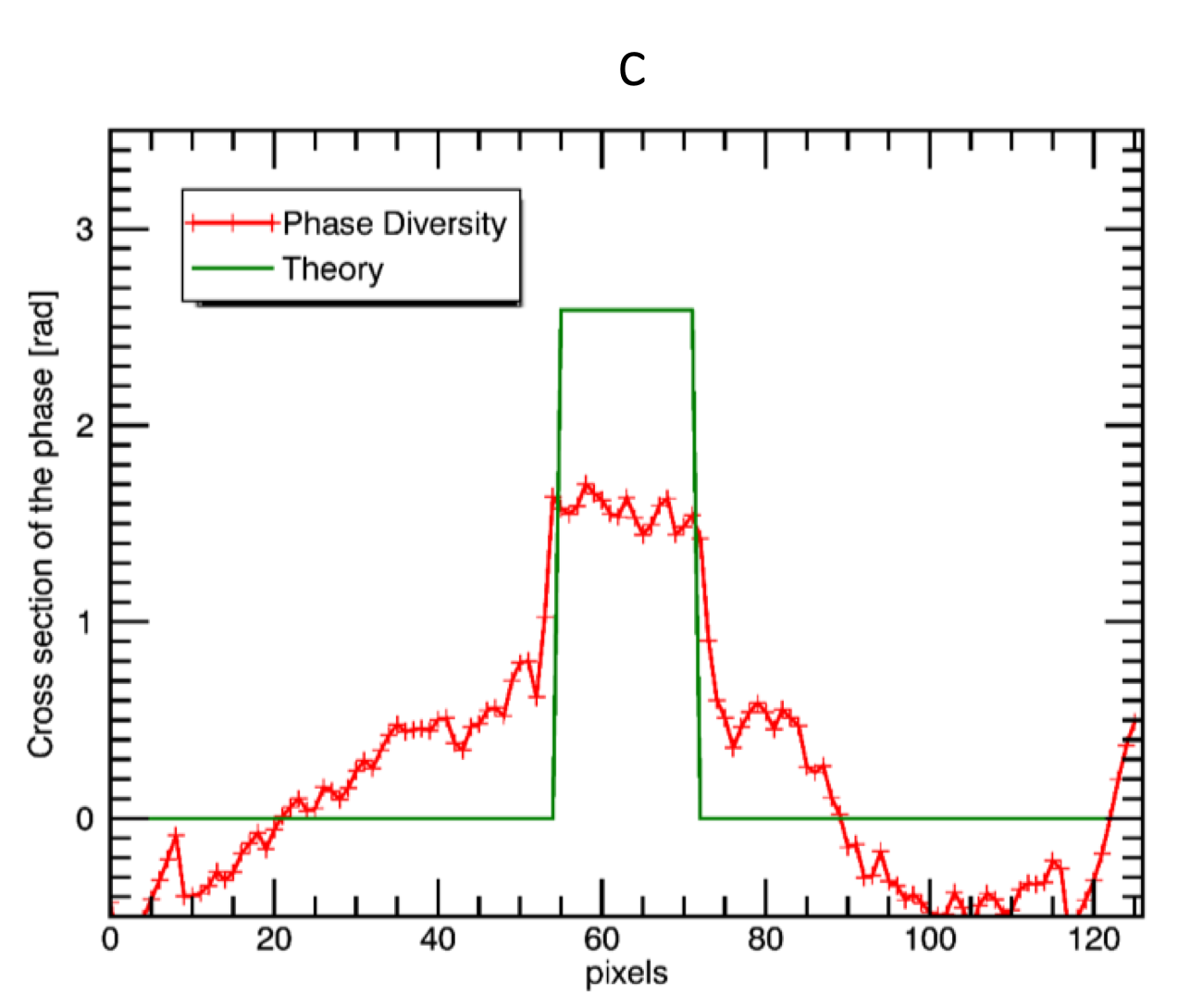}
   \end{tabular}
   \end{center}
   \caption[coron_phase_COFFEE] 
   { \label{fig:coron_phase_COFFEE2} 
Comparison between expected phase and estimated phase when the Iris-AO is flat except for the central segment on which a $225$ nm piston is applied, in presence of the Lyot coronagraph: (a) Theoretical phase map computed from the command sent to the Iris-AO, in radians, (b) Phase map reconstructed by COFFEE, in radians, (c) vertical cross-sections of these two phases. Extra low-order aberrations are reconstructed, in addition to a underestimation of the local piston  on the central segment.}
   \end{figure*}

Figure \ref{fig:coron_phase_COFFEE} shows the phases reconstructed with COFFEE in coronagraphic configuration when different phases are applied on the segments : (a) the Iris-AO mirror is flattened, (b) the Iris-AO mirror is flattened except for the central segment on which a $188$ nm piston is applied, (c) the Iris-AO mirror is flattened except for the central segment in which a $0.2$ mrad tip is applied. On the top line (1) are the raw phases, while on the bottom line (2) the phases are subtracted by the reference phase (1a). On these different phases, the hexagonal pattern can be clearly recognized and on the non reference phases, the moving segment can be identified. However, already at this stage, several issues can be addressed: first, low order aberrations, that should not exist, appear and differ strongly from one estimation to the other. Secondly, the reconstruction amplitude is lower than expected, which is illustrated on Figure \ref{fig:coron_phase_COFFEE2}. On this picture, we can notice that the estimated piston on the central segment is around $150$ nm instead of $263$ nm. These errors of reconstruction are clearly not acceptable and will be studied in the next section. 

\subsection{Impact of the FPM on the coronagraphic image}
\label{s:ImpactOfTheFPMOnTheCoronagraphicImage}

The FPM can explain the errors of phase reconstruction pointed before since it can be seen as a high-pass filter. It filters out the light close to the optical axis, which contains the low-frequency aberrations, such as, in our case, the reference or background phase or even low-frequency components of the segment itself.

\begin{figure*}
   \begin{center}
   \begin{tabular}{ccc}
   \hspace{-0.7cm} \includegraphics[height=3.5cm]{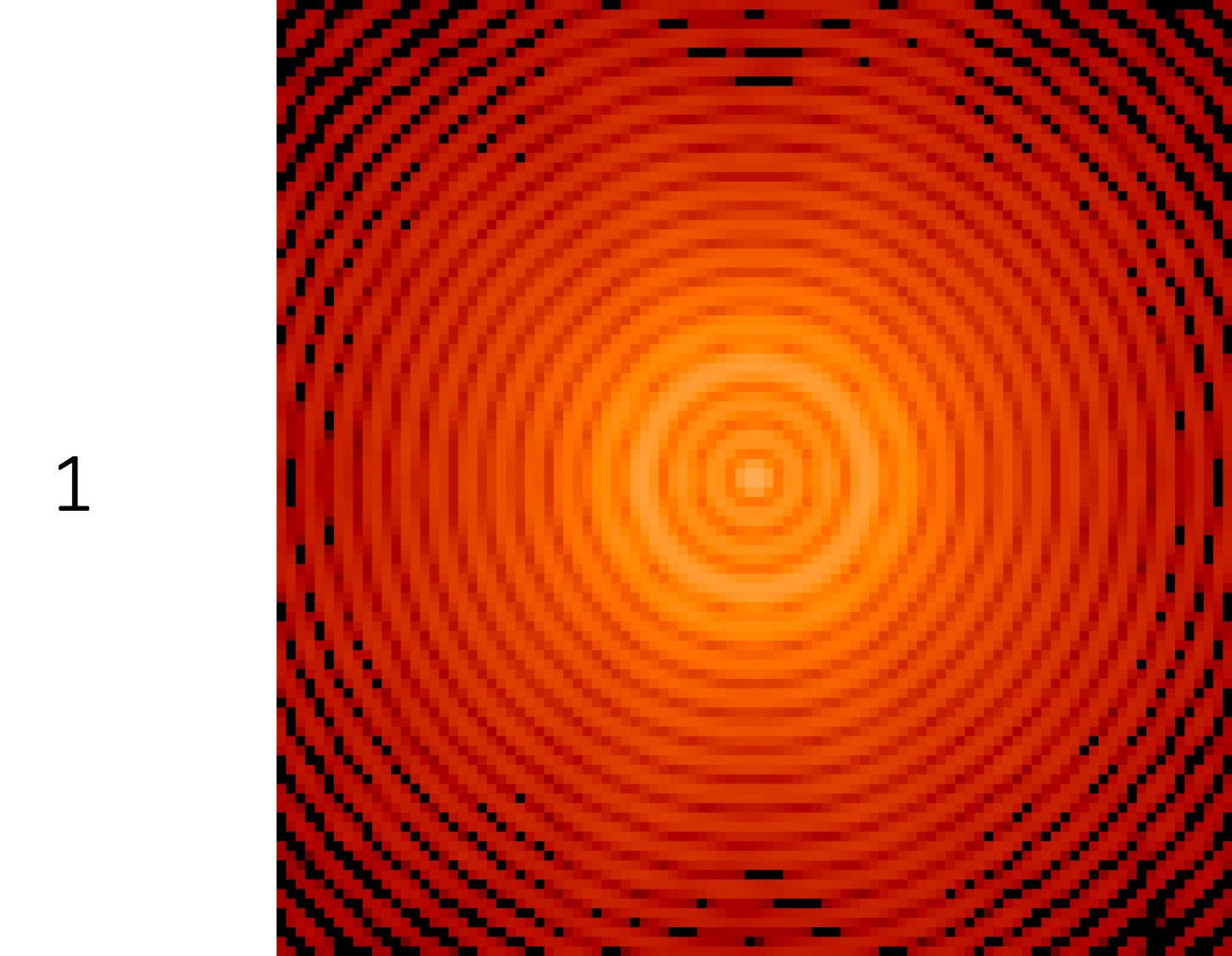} & \includegraphics[height=3.5cm]{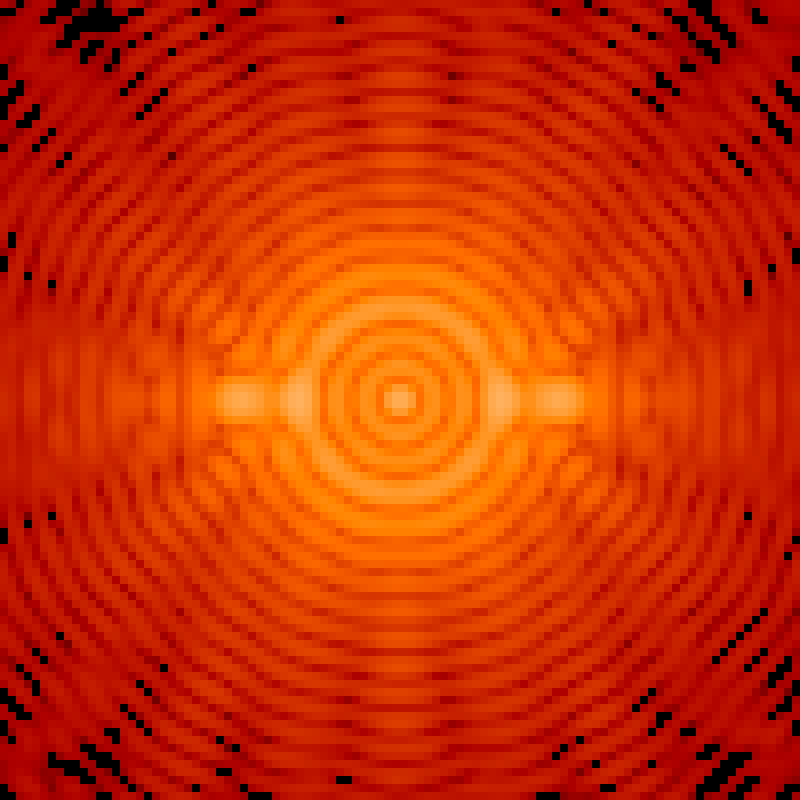} & \includegraphics[height=3.5cm]{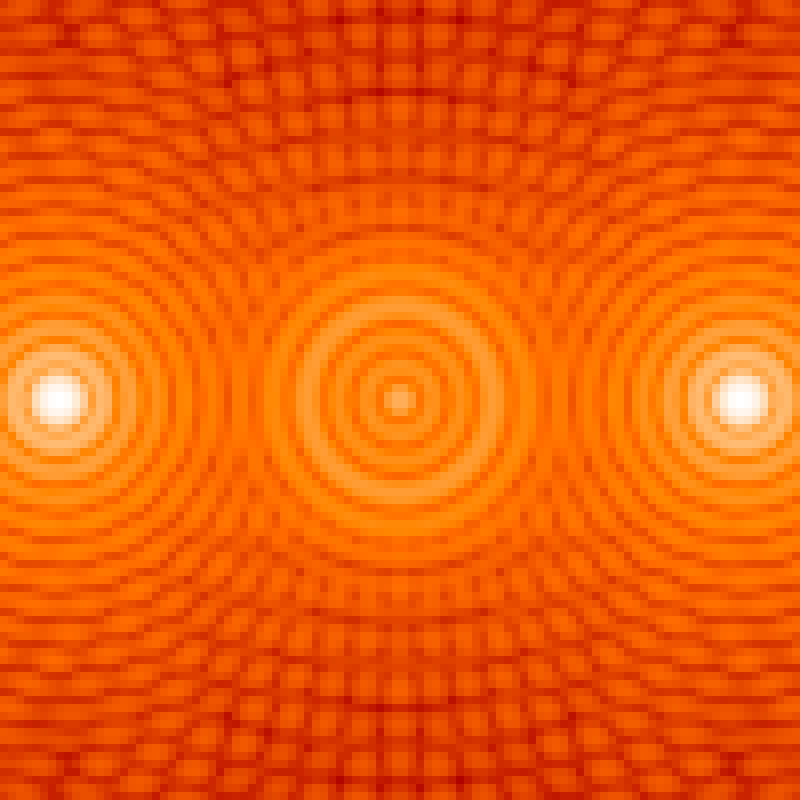} \\
   \hspace{-0.7cm} \includegraphics[height=3.5cm]{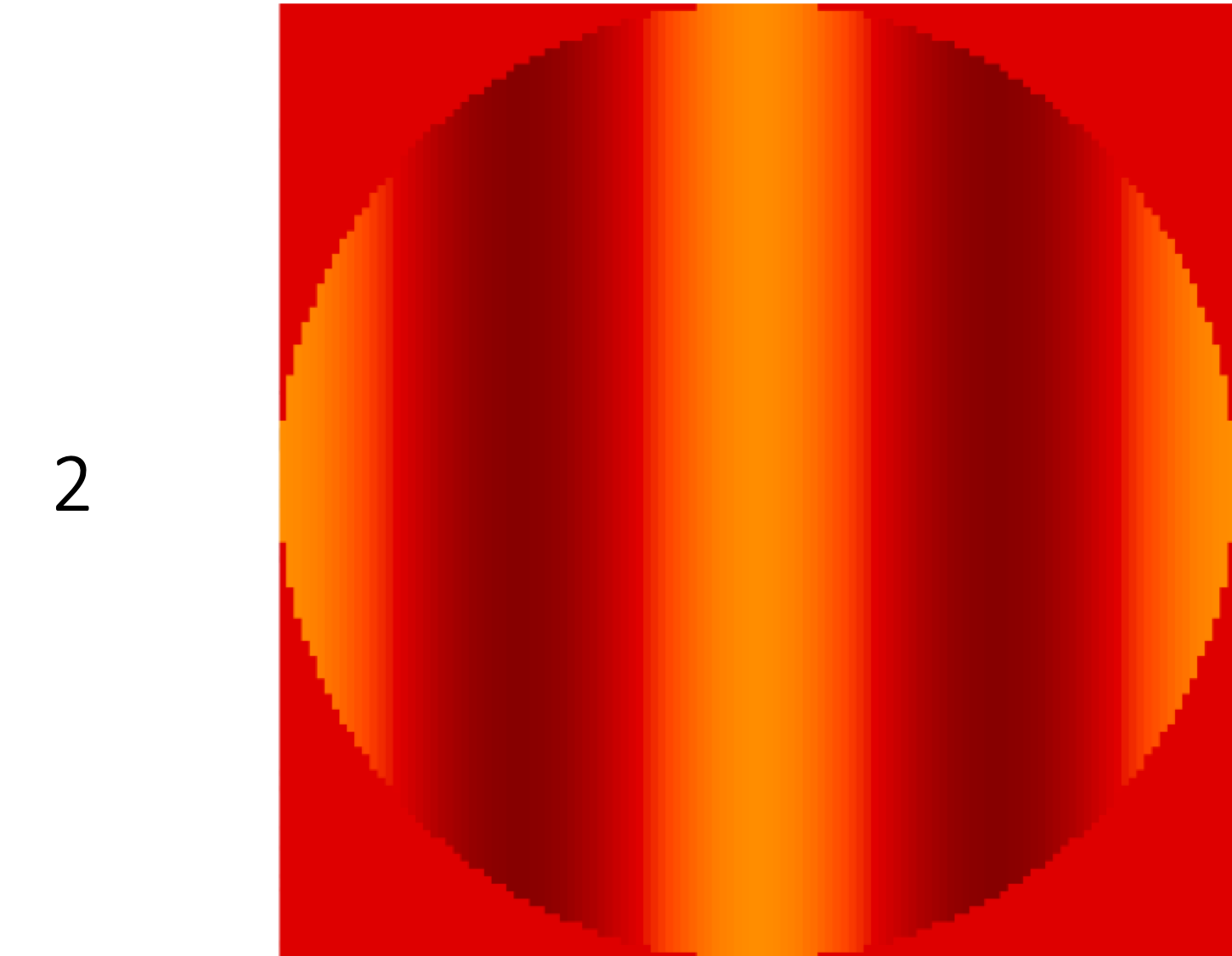} & \includegraphics[height=3.5cm]{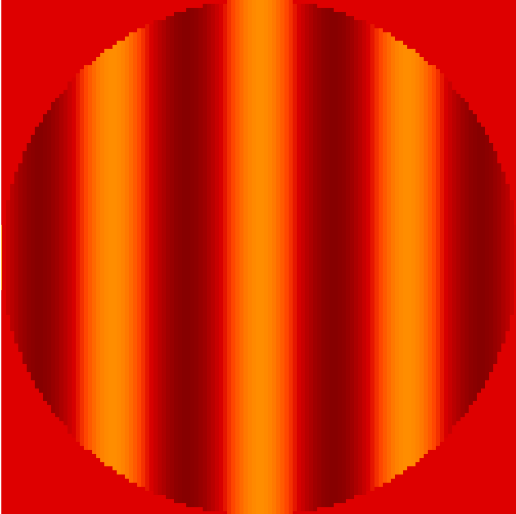} & 
   \includegraphics[height=3.5cm]{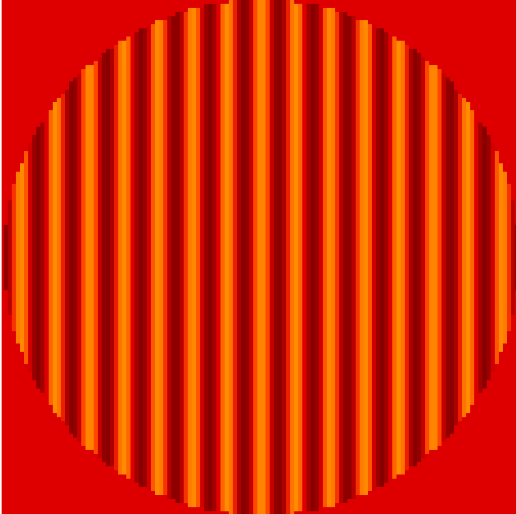} \\
   \hspace{-0.7cm} \includegraphics[height=3.5cm]{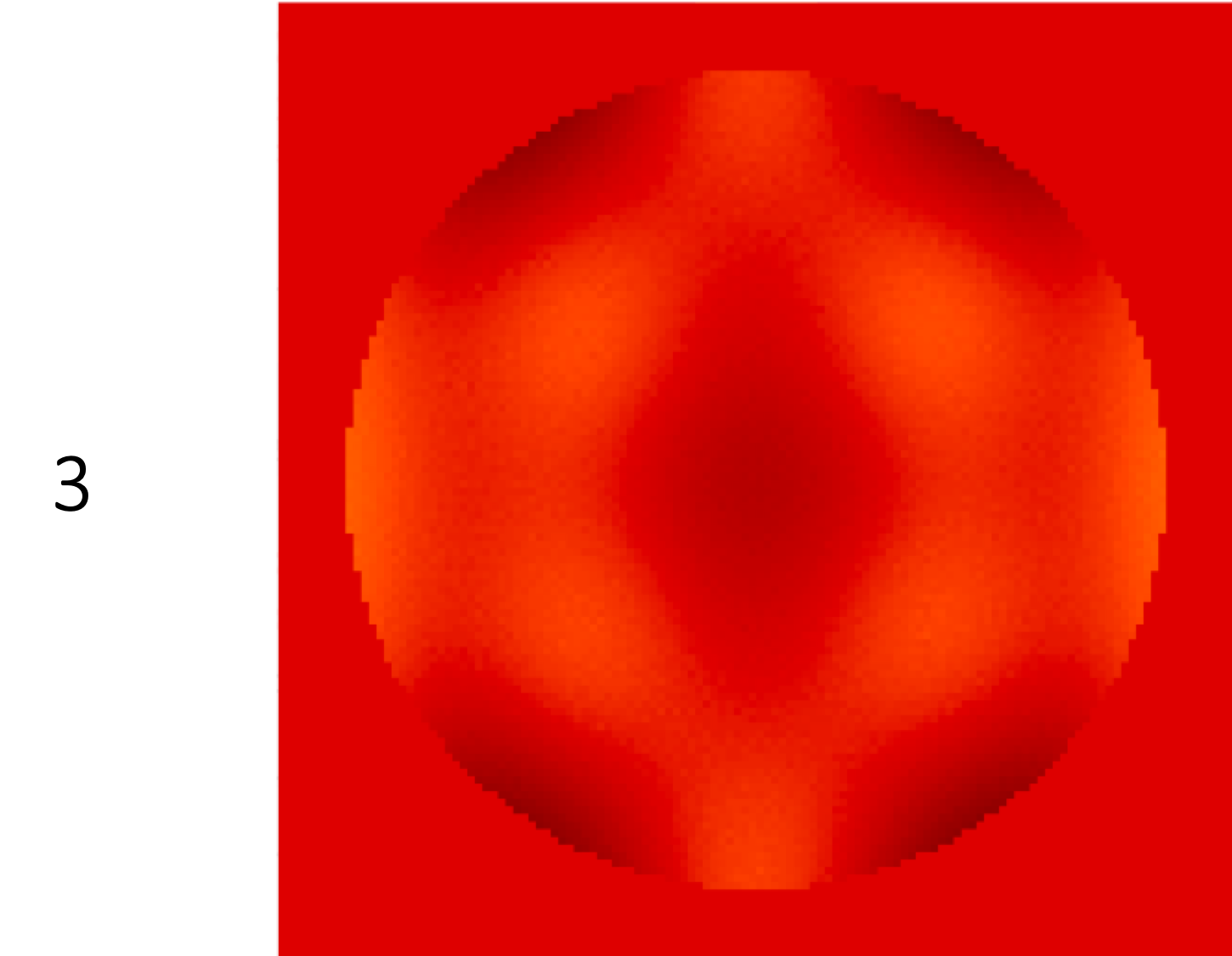} & \includegraphics[height=3.5cm]{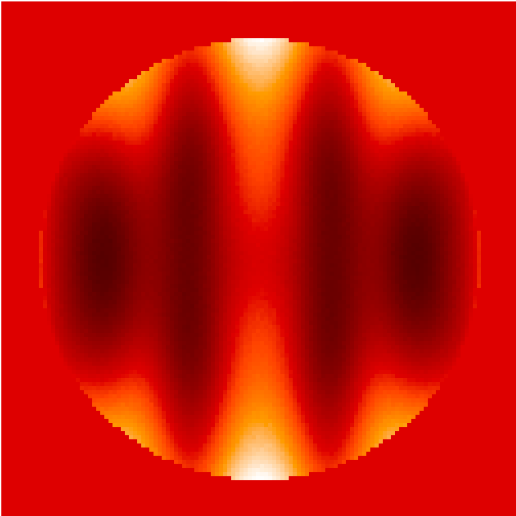} & \includegraphics[height=3.5cm]{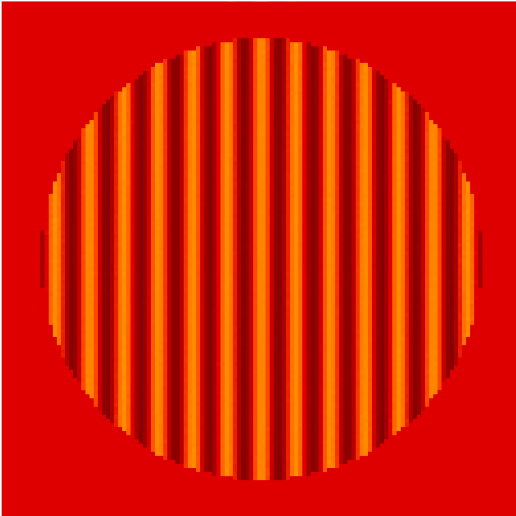} \\
   \hspace{-0.7cm} \includegraphics[height=3.5cm]{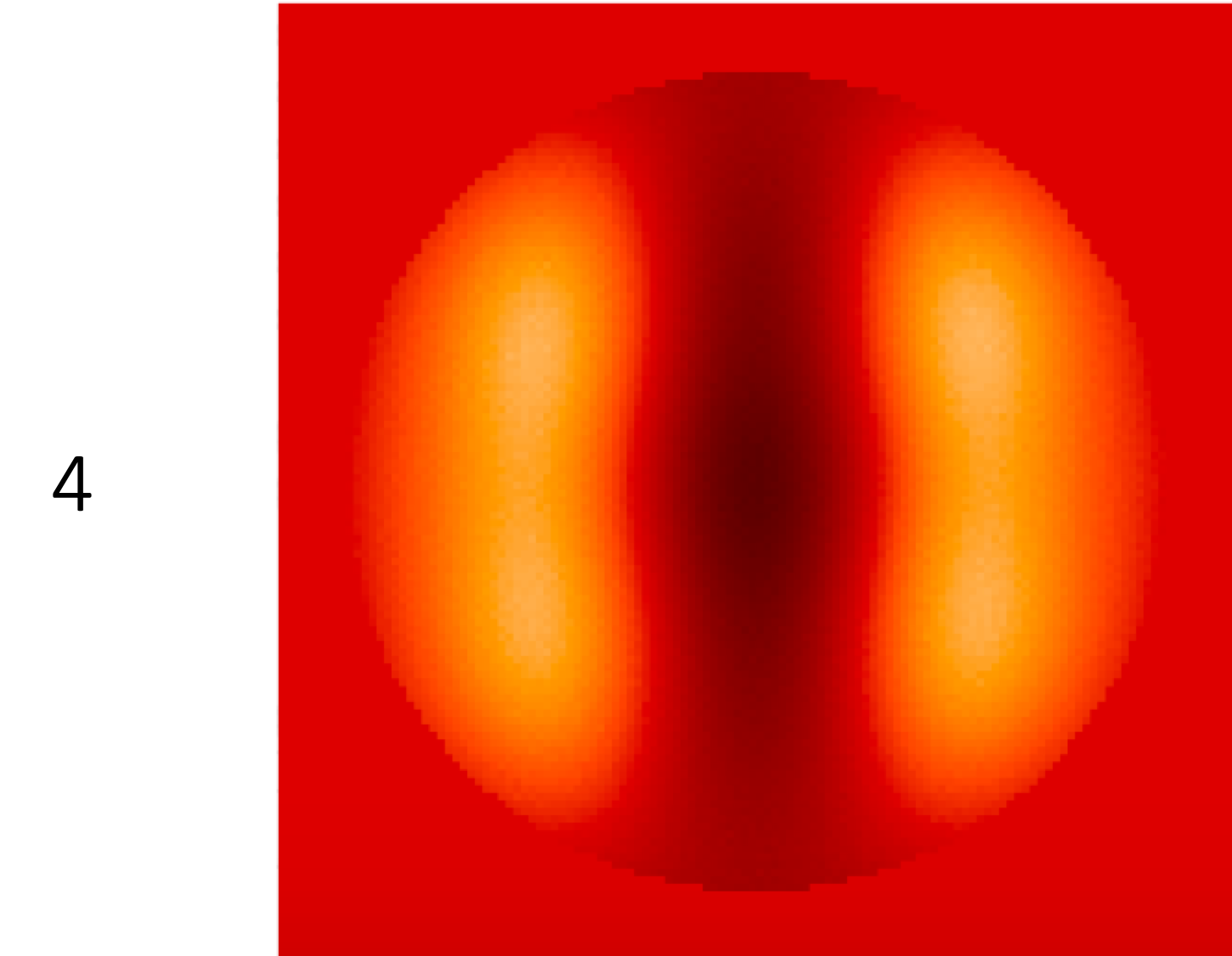} & \includegraphics[height=3.5cm]{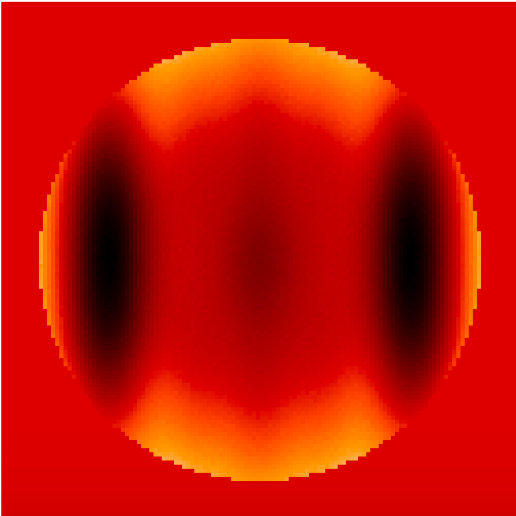} & \includegraphics[height=3.5cm]{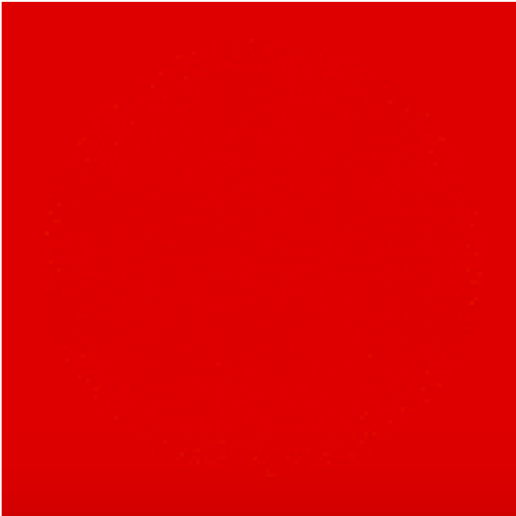} \\
   \includegraphics[height=3.5cm]{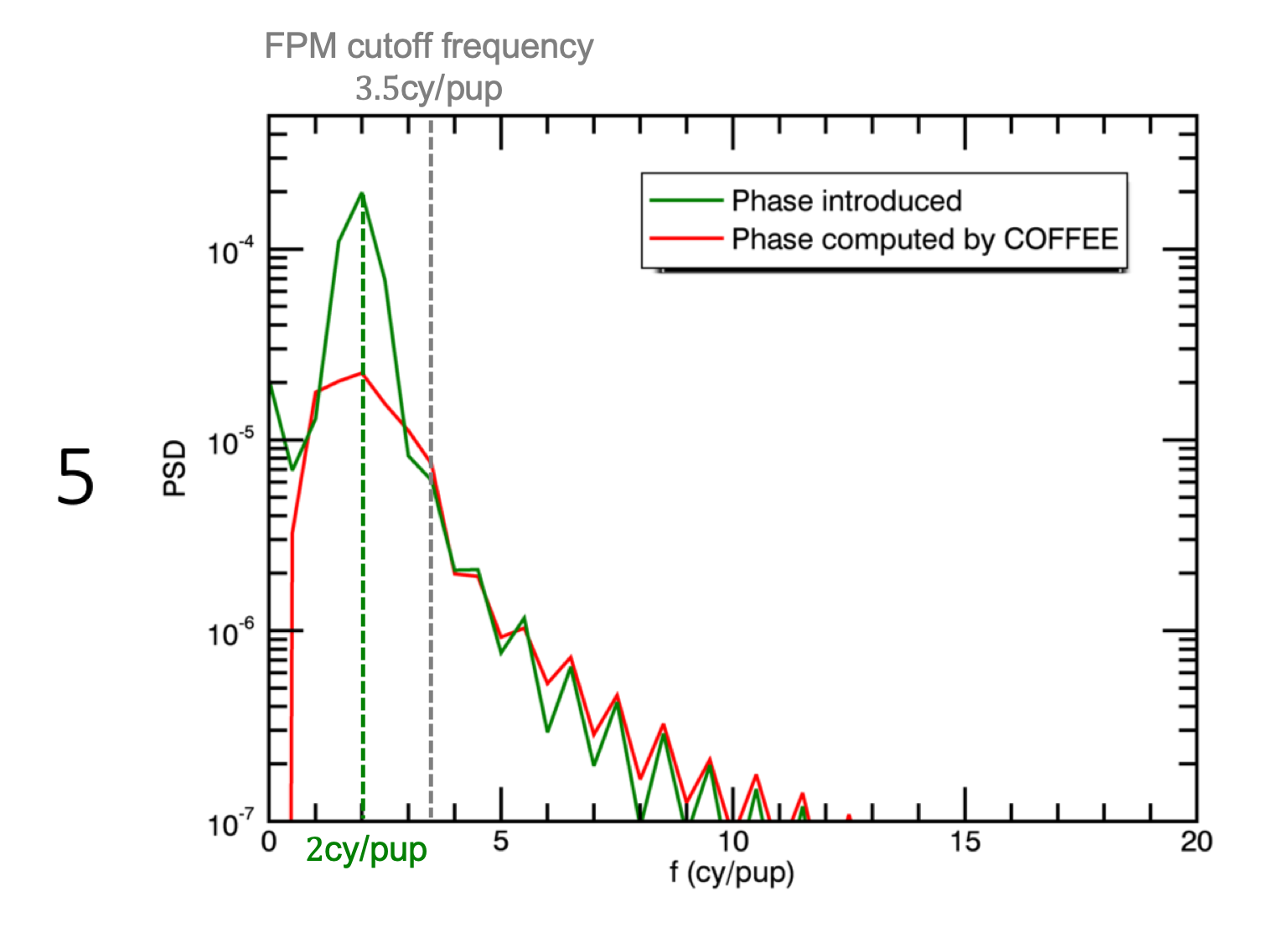} & \includegraphics[height=3.5cm]{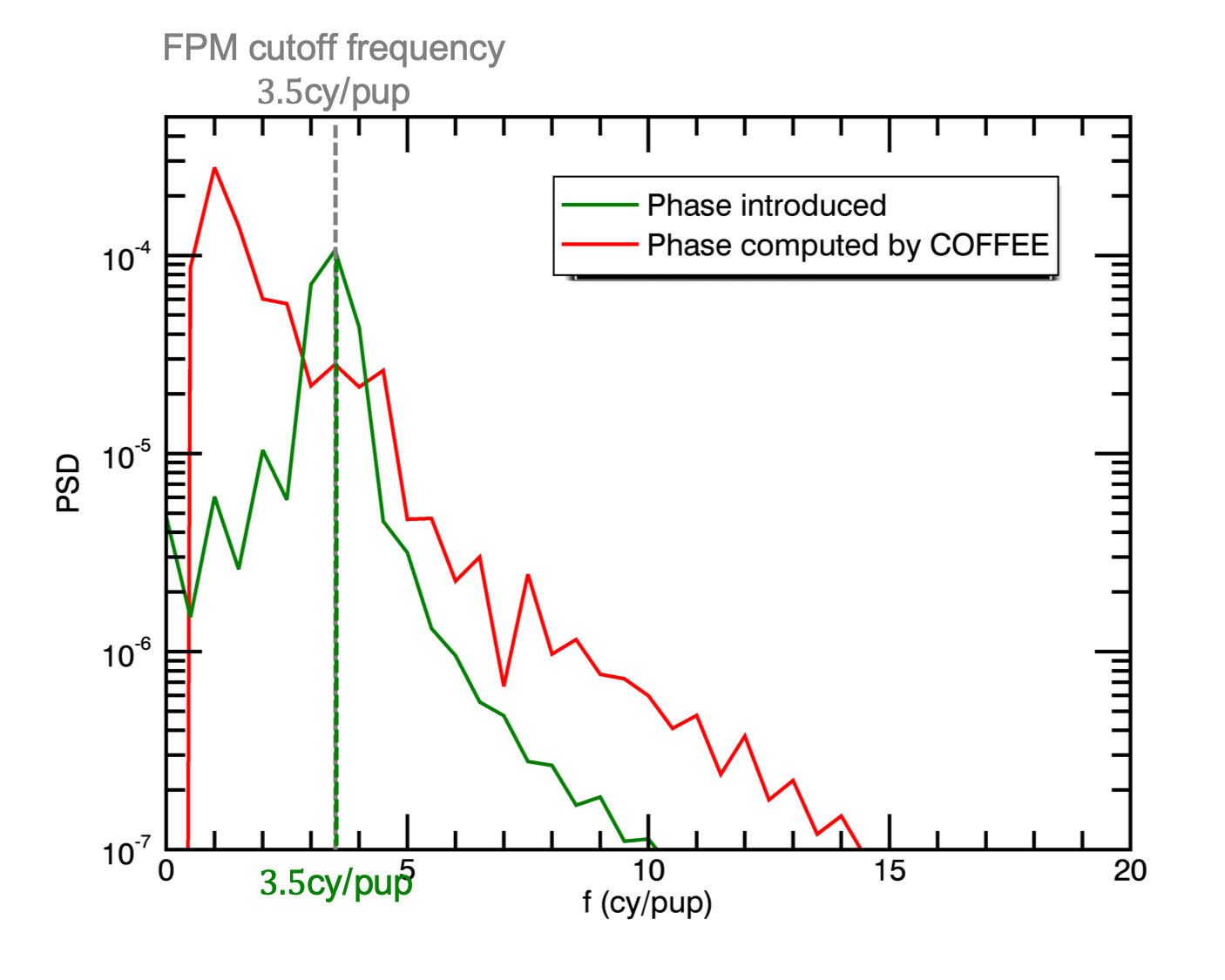} & \includegraphics[height=3.5cm]{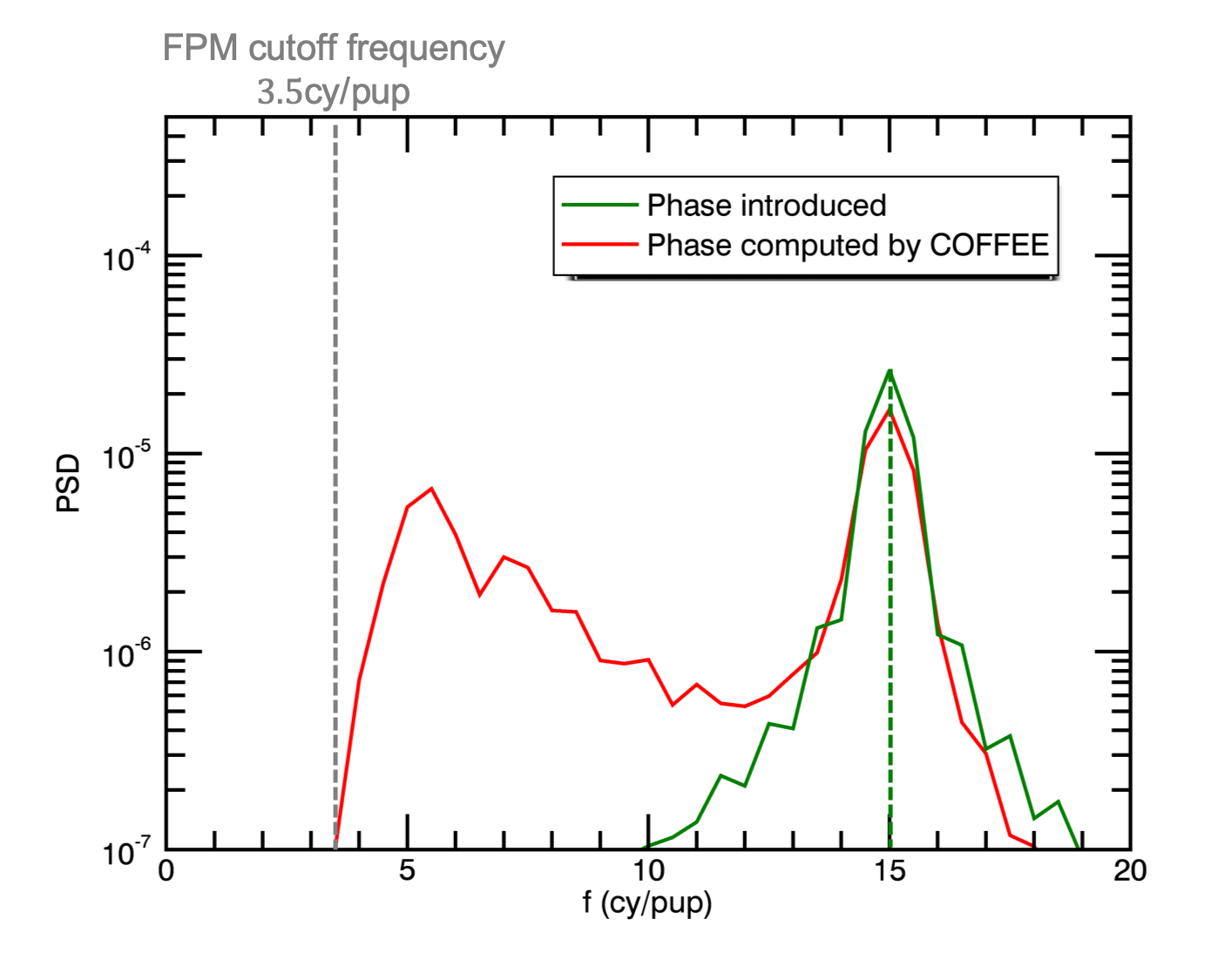} \\
    \hspace{0.4cm} a & b & c
   \end{tabular}
   \end{center}
   \caption[SineWaveCOFFEE] 
   { \label{fig:SineWaveCOFFEE} 
Comparison between theoretical phases and phases reconstructed by COFFEE for sine-like phases with an amplitude of $70$ nm and three different spatial frequencies: (a) two cycles per pupil, (b) 3.5 cycles per pupil, and (c) 15 cycles per pupil. For each case, we show: (1) the PSF in the detector plane, (2) the theoretical phase map to reconstruct, (3) the phase map reconstructed by COFFEE, cropped by the Lyot stop, (4) the difference, cropped by the Lyot stop, between the theoretical phase map and the phase map reconstructed by COFFEE, and (5) the power spectral distribution (PSD) of the theoretical phase and the phase reconstructed by COFFEE. On these curves, the cutoff frequency of the FPM is indicated in grey, and the frequency of introduced sine phase in green.}
   \end{figure*}

To illustrate this phenomenon, we simulate the PSFs (in-focus and out-of-focus) that would be obtained with the HiCAT optical configuration with no aberration except for a sine phase (see Figure \ref{fig:SineWaveCOFFEE}). The sine phase is chosen at different frequencies: two cycles per pupil, which should be too small to pass the FPM, 3.5 cycles per pupil, that should be very close to the border of the FPM, and 15 cycles per pupil, that should fully pass the FPM. All three sine-phases have an amplitude of $70$ nm, and the diversity phase corresponds, like on the HiCAT experiment, to a $300$ nm PV focus. The in-focus PSFs are shown on the first line (1) of Figure \ref{fig:SineWaveCOFFEE}, the corresponding theoretical phases are shown on the second line (2).

These PSFs are used as inputs for COFFEE, which estimates the aberrations. The resulting maps appear at the third line (3) of Figure \ref{fig:SineWaveCOFFEE}, and the difference between these reconstructed phase maps and the theoretical ones are indicated at line (4). In each case, we also plot the power spectral distributions (PSD) of the theoretical and reconstructed phases, visible on the line (5) of Figure \ref{fig:SineWaveCOFFEE}. As a reminder, the PSD corresponds to the square of the Fourier Transform of the phase:
\begin{equation} 
    PSD(\mathbf{u}) = \left \Vert \widehat{\phi_{up}}(\mathbf{u}) \right \Vert ^2,
\end{equation}
where $\widehat{\phi_{up}}$ is the Fourier Transform of $\phi_{up}$. In Figure \ref{fig:SineWaveCOFFEE} (line 4), we plot the azimuthal mean values of the PSD. We can observe that in the three cases, the green peak does correspond to the frequency introduced in the phase. However, only the PSD of the reconstructed phase in the case of 15 cycles per pupil has a well-located peak with the right amplitude. In the two other cases, the peaks have an offset or are under estimated.

In Table \ref{table:SineWaveCOFFEE}, we also indicate the RMS values of the different phases (theoretical, COFFEE-reconstructed, and error, for the total aberrations, the low-frequency components, and the high-frequency components). 

\begin{table*}
\centering
\begin{tabular}{|c|c|c|c|}
  \hline
  Freq (cy/pup) & Theory (rad) & Reconstruction (rad) & Difference (rad) \\
  \hline
  2 & 0.24 & 0.092 & 0.26 \\
  \hline
  3.5 & 0.24 & 0.26 & 0.29 \\
  \hline
  15 & 0.24 & 0.24 & 0.0038 \\
  \hline
\end{tabular}
\caption{RMS wavefront aberrations of the different phases of Figure \ref{fig:SineWaveCOFFEE}, in radians at $\lambda = 638$nm, computed in the Lyot stop only. For each frequency (2, 3.5, and 15 cycles per pupil), the RMS values of the theoretical phase, the phase reconstructed by COFFEE, and the reconstruction error (Diff) are given. The frequencies are indicated in cycles per pupil. We notice that the reconstruction error in the case of 15 cycles per pupil indicates that only this aberration is well reconstructed.}
\label{table:SineWaveCOFFEE}
\end{table*}

\begin{figure}
   \begin{center}
   \includegraphics[height=2.5cm]{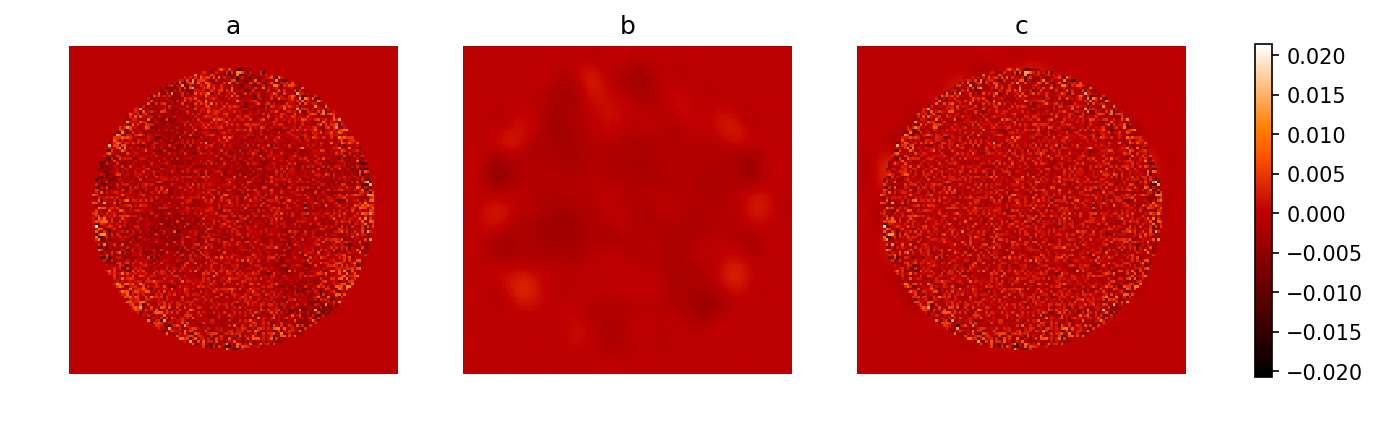}
   \end{center}
   \caption[Diff1] 
   { \label{fig:Diff1} 
Reconstruction error of Figure \ref{fig:SineWaveCOFFEE} (4c). The unit are in radians. (a) Same reconstruction error than Figure \ref{fig:SineWaveCOFFEE} (4c), at an optimized scale. (b) Orders lower than seven cycles per pupil of the reconstruction error (a), obtained with a cut on the PSD and an inverse Fourier Transform. (c) Difference between the phase maps (a) and (b). Patterns of a few cycles per pupil are visible and can be easily identified using a spatial filter. They correspond to the peak at around five cycles per pupil of the PSD of Figure \ref{fig:SineWaveCOFFEE} (5c). It has been shown in Section \ref{s:ResultsWithCorono} that with a coronagraph, the COFFEE algorithm generates extra low-order aberrations.}
   \end{figure}

We can notice that in the high-frequency aberration case (15 cycles per pupil), the phase is well reconstructed, and the RMS of the error is minimal (0.0038 radians RMS). This kind of aberration fully passes the FPM, even if low-order patterns around five cycles/pupil can be seen on the PSD. Since the difference visible on Figure \ref{fig:SineWaveCOFFEE} (4c) not being well visible, it is also shown on Figure \ref{fig:Diff1} at an optimized scale. This reconstruction error is made of numerous spatial frequencies, except for frequencies lower than the size of the FPM (3.5 cycles per pupil). In particular, Figure \ref{fig:Diff1} (b) shows the components with frequencies lower than seven cycles per pupil. This cutoff frequency has been chosen to identify the peak at around five cycles per pupil visible in the PSD of Figure \ref{fig:SineWaveCOFFEE} (5c). It has been shown in Section \ref{s:ResultsWithCorono} that with a FPM, the COFFEE algorithm reconstructs low-order aberrations that should be absent, this phenomenon is also visible on the PSD of Figure \ref{fig:SineWaveCOFFEE} (5b).


Conversely, the reconstruction is not efficient for low-order aberrations such as the two cycles per pupil case. The reconstructed phase is very different than the theoretical one, in terms of phase pattern and RMS value (0.29 radians RMS), most of the error coming from low-frequency aberrations (see Table \ref{table:SineWaveCOFFEE}).

This test indicates that there exist some modes that are not or badly visible by COFFEE, mainly when they are blocked by the FPM. The FPM limits COFFEE's ability to sense and consequently reconstruct low-order modes with spatial frequencies smaller than the spatial extent of the FPM, that is, smaller than $3.5$ cy/pup in our case (as a reminder, the FPM has a diameter of $7.2 \pm 0.9 \lambda/D$). On the other hand, COFFEE shows its full capability for modes with higher spatial frequencies than the size of the FPM.

\subsection{Impact of segment-level aberrations on the phase reconstruction}
\label{s:ImpactOfSegmentLevelAberrationsOnThePhaseReconstruction}

From the previous study, we can deduce that phase patterns containing low-order variations are not properly reconstructed. Given that segment-level aberrations contain low-frequency components, COFFEE cannot fully reconstruct phasing errors, and can even add low-order background phase aberrations.

\begin{figure*}
   \begin{center}
   \begin{tabular}{c}
   \includegraphics[height=5cm]{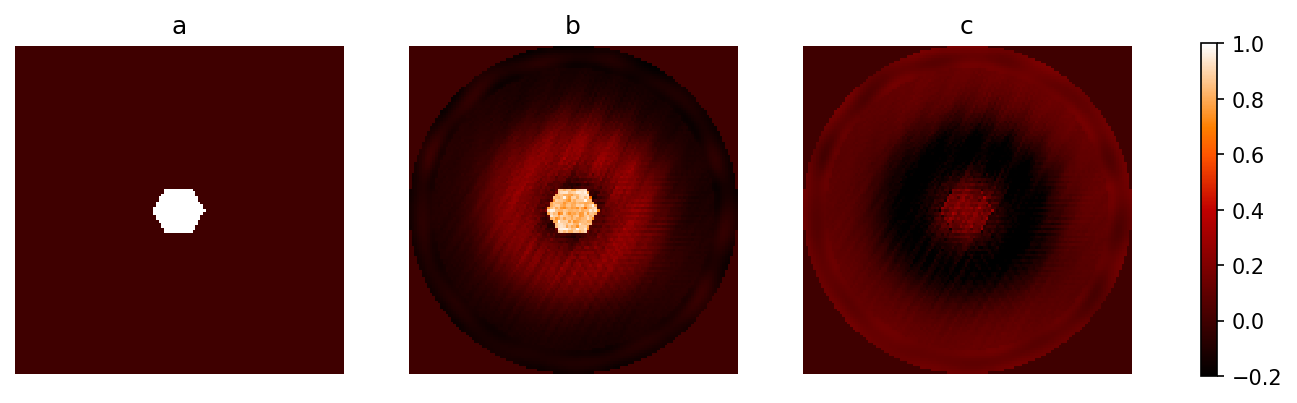} \\
   \includegraphics[height=7cm]{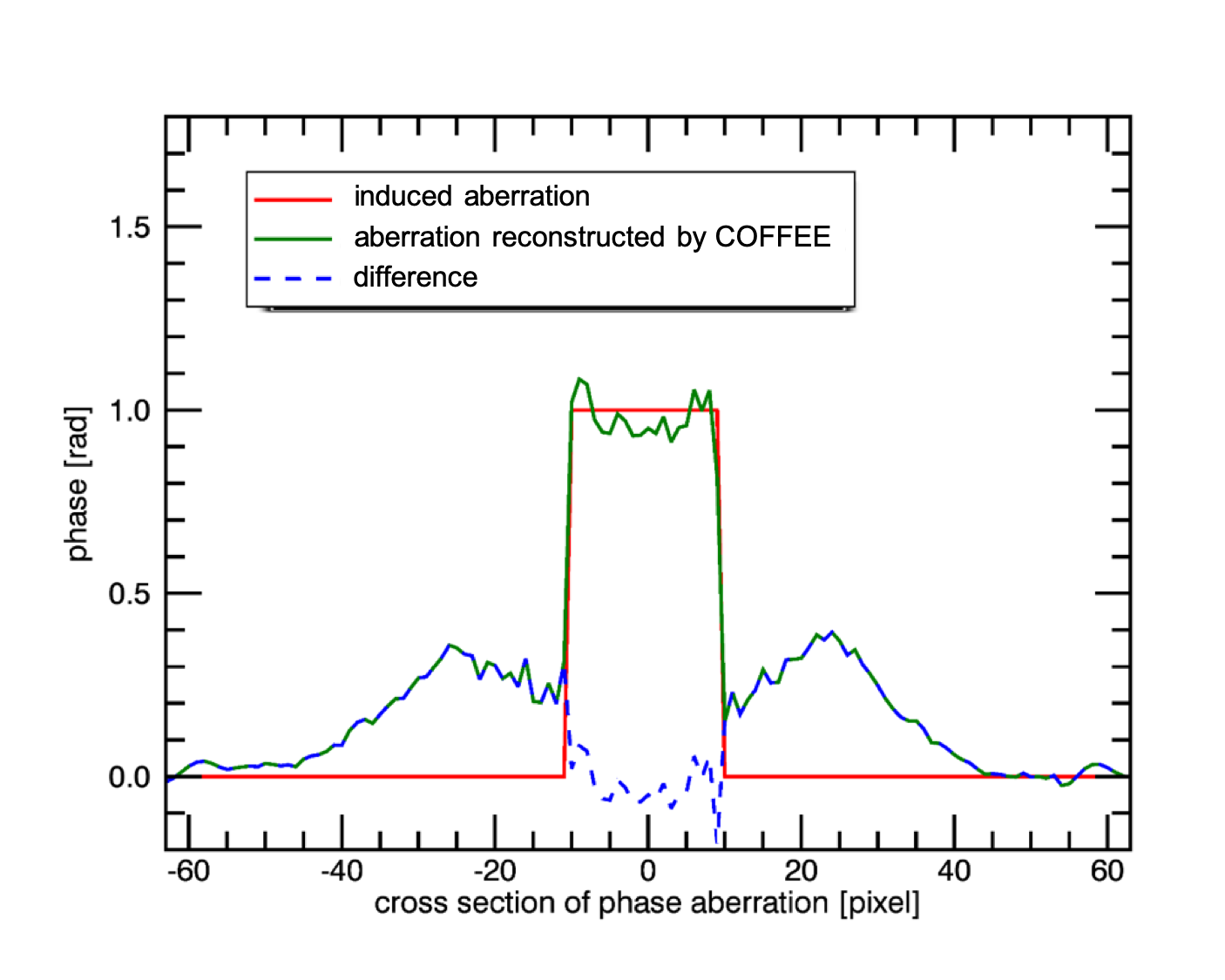}
   \end{tabular}
   \end{center}
   \caption[simu_1seg] 
   { \label{fig:simu_1seg} 
Test of reconstruction of a segment-level phase error with COFFEE in simulation only: (a) theoretical phase to reconstruct: a 1 rad piston is applied on the central segment, (b) phase reconstructed with COFFEE, (c) difference between the two previous phases, (d) cross-sections of the three phases of Figure \ref{fig:simu_1seg} along the vertical axis.. We can notice in particular the reconstruction of a low-order aberration that is not present in the theoretical phase.}
   \end{figure*}

To theoretically investigate this effect, we reconstruct a segment-level piston error with COFFEE. To do so, we simulate in- and out-of-focus PSFs with a segment-level phase error (piston of $1$ rad PV on the central segment) in the same configuration as the HiCAT Iris-AO. These PSFs are used as inputs of COFFEE to reconstruct the phase error. The results are shown in Figure \ref{fig:simu_1seg}.

\begin{figure}
   \begin{center}
   \begin{tabular}{c}
   \includegraphics[width=8.5cm]{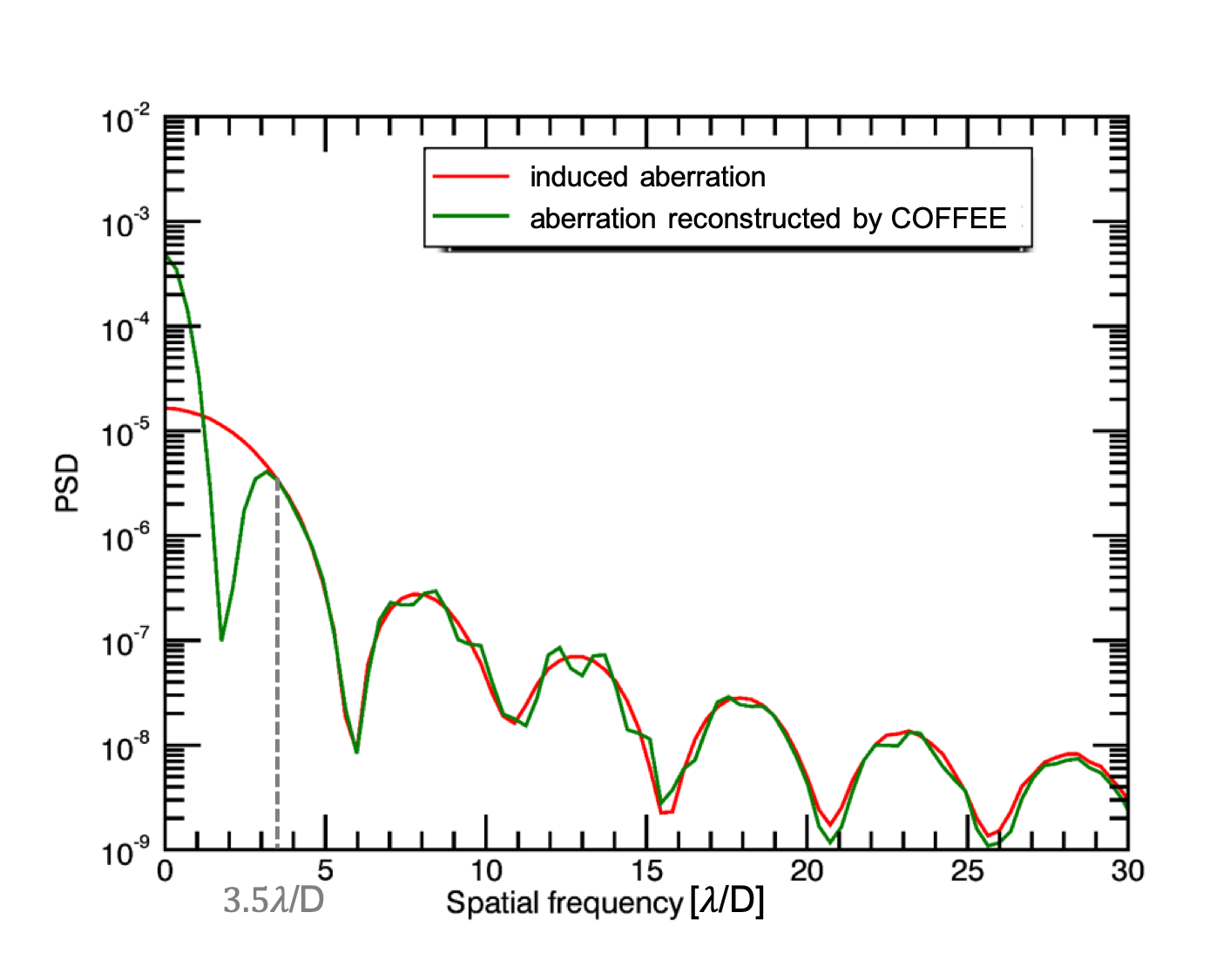}
   \end{tabular}
   \end{center}
   \caption[simu_1seg_DSP] 
   { \label{fig:simu_1seg_DSP} 
(red) Power spectral distribution (PSD) of a flat phase with one segment unphased, at the same scale and dimensions as the Iris-AO seen by COFFEE. (green) PSD of the corresponding phase reconstructed by COFFEE. We observe a fit for frequencies higher than the FPM cutoff frequency (around $3.5 \lambda/D$), while lower-frequency components are not well reconstructed.}
   \end{figure}

Figure \ref{fig:simu_1seg_DSP} also indicates in red the theoretical PSD of a flat phase, except for the central segment. The PSD is higher below 6-7 cy/pup, that is, spatial frequencies that are mostly cut by the FPM. The PSD of the reconstructed phase is also plotted in green on Figure \ref{fig:simu_1seg_DSP} and can be compared to the theoretical one. It can be noted that above the FPM cut-off frequency, the reconstructed frequencies fit the theoretical ones, while at low frequencies, the errors of reconstruction drastically increase.

\begin{figure*}
   \begin{center}
   \begin{tabular}{c}
   \includegraphics[height=4cm]{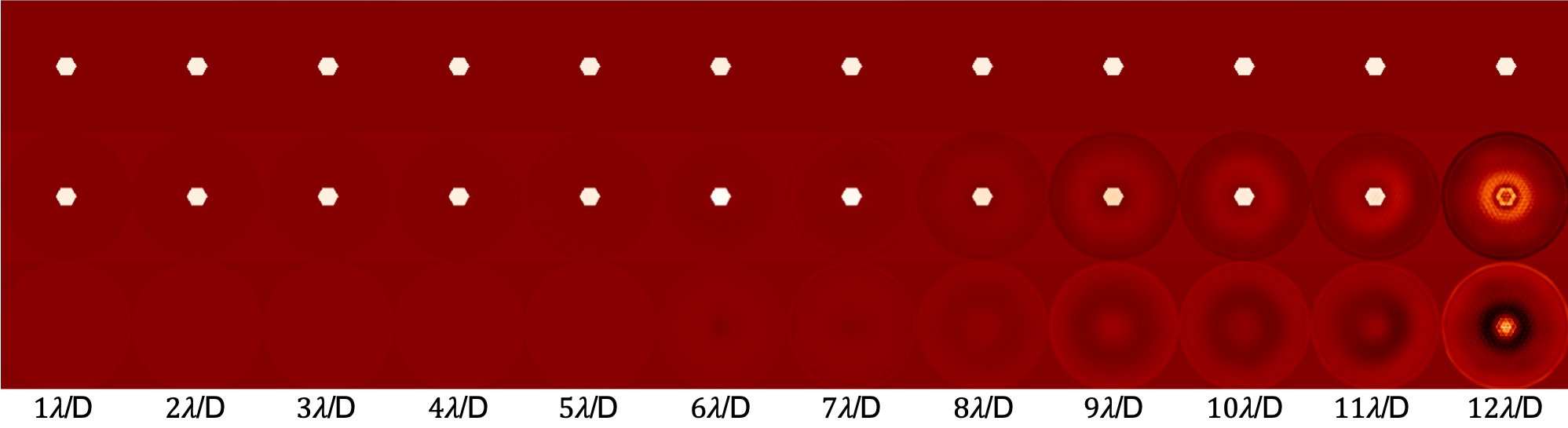}
   \end{tabular}
   \end{center}
   \caption[simu_1seg_LS90percent_FPM1a12lambdaD] 
   { \label{fig:simu_1seg_LS90percent_FPM1a12lambdaD} 
Test of different FPM sizes and impact on the reconstruction : (top) theoretical phases, (middle) phases reconstructed with COFFEE, (bottom) difference between the two previous phases, from left to right : $1$ to $12$ $\lambda/D$, for a Lyot ratio of $90\%$. After a diameter of around $7 \lambda/D$, the FPM acts like a high-pass filter and blocks low frequency components of the segment-like aberration that cannot be seen by the sensor.}
   \end{figure*}

Therefore, COFFEE shows difficulties to efficiently estimate phases containing the frequencies cut by the FPM. We may then wonder how the size of the FPM impacts the reconstruction of low-order aberrations, in particular segment-level aberrations. Figure \ref{fig:simu_1seg_LS90percent_FPM1a12lambdaD} presents the theoretical (top), reconstructed (middle), and difference (bottom) phases, for a FPM with a radius from $1$ to $12$ $\lambda/D$ (left to right) and a Lyot ratio of $90\%$. We can notice a clear deterioration of the reconstruction phases when the FPM size increases, particularly above $7 \lambda/D$. This can be explained by the fact that in our case we have around seven segments along the pupil diameter. As a consequence, the most information in the focal plane about each segment is contained in a circle of diameter around $7 \lambda/D$ (FWHM of the Fourier Transform of one segment). Therefore, a FPM smaller than this diameter lets more information pass, while a larger FPM will cut most of the segment information.

We can conclude here that the size of the FPM has a noticeable impact on the reconstruction of segment-level aberrations, and that the reconstruction error of segment-level aberrations on the HiCAT testbed could probably be mitigated with a smaller FPM. Using a smaller FPM is also a scientific objective, since it enables to detect companions such as Earth-like planets closer to the host star.


\section{Conclusions}
\label{s:Conclusions}

In this article, we carry out an experimental test of COFFEE on the HiCAT testbed to reconstruct segment phasing errors. It has been tested in two different configurations: without and with the coronagraph.

Without a coronagraph, COFFEE becomes equivalent to a traditional phase diversity algorithm. The reconstruction of piston, tip, and tilt aberrations was linear between $0$ and $200$ nm.

With a coronagraph, COFFEE was able to identify and analyze an intrinsic limitation of focal plane wavefront sensors in presence of a Lyot coronagraph: the FPM behaves like a high-pass filter and blocks low spatial frequencies that are present in segment-level aberrations. Therefore this sensor cannot reconstruct frequencies lower than its cut-off frequency: it cannot fully estimate low-order aberrations and it may even add absent low-order aberrations.

\begin{figure}
   \begin{center}
   \begin{tabular}{c}
   \includegraphics[width=8.5cm]{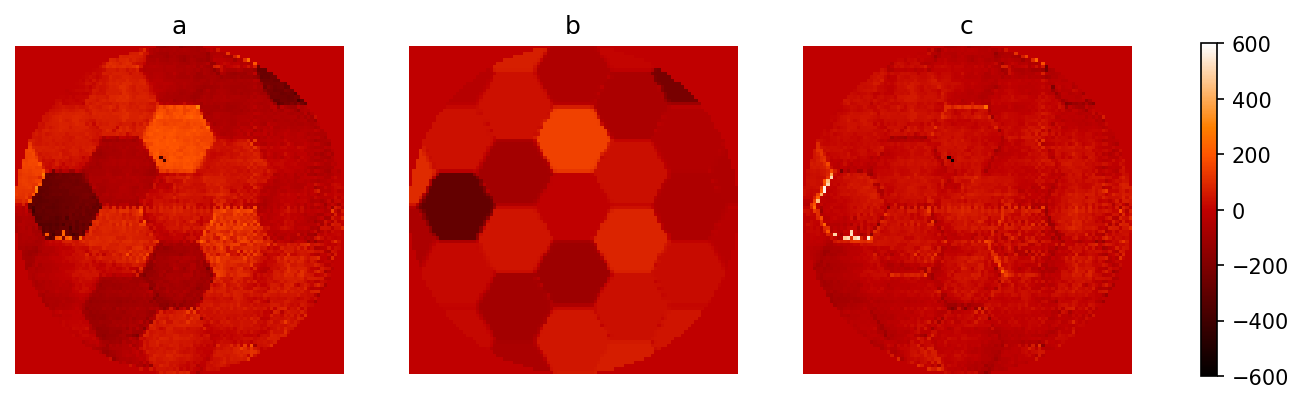}
   \end{tabular}
   \end{center}
   \caption[Mars_phases2] 
   { \label{fig:Mars_phases2} 
Comparison, in the coronagraphic configuration (FPM diameter: $4.4 \lambda/D$), of (left) the phase reconstructed by COFFEE subtracted by a reference phase (in nanometers), also reconstructed by COFFEE ($95.0$ nm RMS), (center) the theoretical phase (in nanometers) issued from the command sent to the Iris-AO ($92.3$ nm RMS), and (c) the reconstruction error phase (in nanometers), that is, the difference between the two previous phases, when random piston values are applied on the Iris-AO. The three phases are on the same scale.}
   \end{figure}

A first simple solution to this limitation would be to use a smaller coronagraph, but it would also mean enabling more low-order signal to reach the detector. To test this, we conducted a fast experiment on HiCAT with a $4.4 \lambda/D$-diameter FPM set up on the testbed. Figure \ref{fig:Mars_phases2} shows the reconstructed, command, and difference phases when random pistons with an RMS of $92.3$ nm are applied on the Iris-AO. However, such a solution seems naive since it enters into contradiction with current optimizations of coronagraph design, which tend to have a large FPM to block as much starlight as possible. 

Another approach to mitigate that effect is to modify not the coronagraph size but the density of segments: smaller segments have a different PSD with higher order components. With a telescope like the LUVOIR proposition, which has 120 segments \citep{LUVOIR2019}, COFFEE would be more efficient since more segment spatial frequencies would pass the coronagraph.

So far, only hardware solutions have been proposed, but an optimization on the COFFEE parameters could perhaps be implemented. In particular, the regularization of the COFFEE optimizer could be better better in fitting segment-like aberrations. \cite{HerscoviciSchiller2018b} also encountered unsensed modes when they applied the COFFEE sensor on the THD2 testbed, with a four quadrant phase mask coronagraph. They mitigated this limitation by setting up a optimized regularization inside the algorithm itself. This numerical solution requires more investigation in the case of a Lyot coronagraph, since the low-order modes, cut by the FPM, also specifically correspond to the ones we are interested in: they are main spatial components of segment-like aberrations.

Eventually a filter could be added in COFFEE to restrict the reconstruction on high-order aberrations. Low-order aberrations can be reconstructed by a separate branch set up with a specific tool such as a low-order wavefront sensor (LOWFS). Most future instrument concepts, including the HiCAT testbed, include a LOWFS, which could be combined with a COFFEE-like sensor to fully reconstruct aberrations and in particular phasing errors.

\begin{acknowledgements}
This work was supported in part by the National Aeronautics and Space Administration under Grant 80NSSC19K0120 issued through the Strategic Astrophysics Technology / Technology Demonstration for Exoplanet Missions Program (SAT-TDEM; PI:  R. Soummer).

This work has also received support of IRIS Origines et Conditions d’Apparition de la Vie (OCAV) of PSL Idex under the program « Investissements d’Avenir » with the reference ANR-10-IDEX-0001-02 PSL.

The author would also like to thank the LESIA laboratory (Laboratoire d'Etudes Spatiales et d'Instrumentation en Astrophysique) of the Observatoire de Paris, the French Aerospace Lab (ONERA) in the frame of the VASCO Research Project and the Laboratoire d'Astrophysique de Marseille (LAM), and the ANR WOLF for their support.

The author particularly thanks Rapha\"el Galicher for his help and support.
\end{acknowledgements}

\bibliographystyle{aa} 
\bibliography{bib}

\begin{thebibliography}{54}
\expandafter\ifx\csname natexlab\endcsname\relax\def\natexlab#1{#1}\fi

\bibitem[{{Baudoz} {et~al.}(2006){Baudoz}, {Boccaletti}, {Baudrand}, \&
  {Rouan}}]{Baudoz2006}
{Baudoz}, P., {Boccaletti}, A., {Baudrand}, J., \& {Rouan}, D. 2006, in IAU
  Colloq. 200: Direct Imaging of Exoplanets: Science Techniques, ed. C.~{Aime}
  \& F.~{Vakili}, 553--558

\bibitem[{{Brady} {et~al.}(2018){Brady}, {Moriarty}, {Petrone}, {Laginja},
  {Brooks}, {Comeau}, {Leboulleux}, \& {Soummer}}]{Brady2018}
{Brady}, G.~R., {Moriarty}, C., {Petrone}, P., {et~al.} 2018, in

\bibitem[{{Brown} \& {Soummer}(2010)}]{Brown2010}
{Brown}, R.~A. \& {Soummer}, R. 2010, \apj, 715, 122

\bibitem[{{Dalcanton} {et~al.}(2015){Dalcanton}, {Seager}, {Aigrain}, {Battel},
  {Brandt}, {Conroy}, {Feinberg}, {Gezari}, {Guyon}, {Harris}, {Hirata},
  {Mather}, {Postman}, {Redding}, {Schiminovich}, {Stahl}, \&
  {Tumlinson}}]{Dalcanton2015}
{Dalcanton}, J., {Seager}, S., {Aigrain}, S., {et~al.} 2015, ArXiv e-prints
  [\eprint[arXiv]{1507.04779}]

\bibitem[{{Davies} {et~al.}(2010){Davies}, {Ageorges}, {Barl}, {Bedin},
  {Bender}, {Bernardi}, {Chapron}, {Clenet}, {Deep}, {Deul}, {Drost},
  {Eisenhauer}, {Falomo}, {Fiorentino}, {F{\"o}rster Schreiber}, {Gendron},
  {Genzel}, {Gratadour}, {Greggio}, {Grupp}, {Held}, {Herbst}, {Hess},
  {Hubert}, {Jahnke}, {Kuijken}, {Lutz}, {Magrin}, {Muschielok}, {Navarro},
  {Noyola}, {Paumard}, {Piotto}, {Ragazzoni}, {Renzini}, {Rousset}, {Rix},
  {Saglia}, {Tacconi}, {Thiel}, {Tolstoy}, {Trippe}, {Tromp}, {Valentijn},
  {Verdoes Kleijn}, \& {Wegner}}]{Davies2010}
{Davies}, R., {Ageorges}, N., {Barl}, L., {et~al.} 2010, in \procspie, Vol.
  7735, Ground-based and Airborne Instrumentation for Astronomy III, 77352A

\bibitem[{{Dohlen} {et~al.}(2006){Dohlen}, {Langlois}, {Lanzoni}, {Mazzanti},
  {Vigan}, {Montoya}, {Hernand ez}, {Reyes}, {Surdej}, \&
  {Yaitskova}}]{Dohlen2006}
{Dohlen}, K., {Langlois}, M., {Lanzoni}, P., {et~al.} 2006, Society of
  Photo-Optical Instrumentation Engineers (SPIE) Conference Series, Vol. 6267,
  {ZEUS: a cophasing sensor based on the Zernike phase contrast method}, 626734

\bibitem[{{Fauvarque} {et~al.}(2016){Fauvarque}, {Neichel}, {Fusco}, {Sauvage},
  \& {Girault}}]{Fauvarque2016}
{Fauvarque}, O., {Neichel}, B., {Fusco}, T., {Sauvage}, J.-F., \& {Girault}, O.
  2016, Optica, 3, 1440

\bibitem[{{Gonsalves}(1982)}]{Gonsalves1982}
{Gonsalves}, R.~A. 1982, Optical Engineering, 21, 829

\bibitem[{{Harness}(2016)}]{Harness2016}
{Harness}, A.~D. 2016, PhD thesis, University of Colorado at Boulder

\bibitem[{{Herscovici-Schiller}
  {et~al.}(2018{\natexlab{a}}){Herscovici-Schiller}, {Mugnier}, {Baudoz},
  {Galicher}, {Sauvage}, {Patru}, {Leboulleux}, {Vigan}, {Dohlen}, {Fusco},
  {Pueyo}, {Soummer}, {Le Duigou}, {Shiri}, {Sivaramakrishnan}, {St. Laurent},
  {Valenzuela}, \& {Zimmerman}}]{HerscoviciSchiller2018}
{Herscovici-Schiller}, O., {Mugnier}, L.~M., {Baudoz}, P., {et~al.}
  2018{\natexlab{a}}, in

\bibitem[{{Herscovici-Schiller}
  {et~al.}(2018{\natexlab{b}}){Herscovici-Schiller}, {Mugnier}, {Baudoz},
  {Galicher}, {Sauvage}, \& {Paul}}]{HerscoviciSchiller2018b}
{Herscovici-Schiller}, O., {Mugnier}, L.~M., {Baudoz}, P., {et~al.}
  2018{\natexlab{b}}, \aap, 614, A142

\bibitem[{Herscovici-Schiller {et~al.}(2019)Herscovici-Schiller, Sauvage,
  Mugnier, Dohlen, \& Vigan}]{Herscovici2019}
Herscovici-Schiller, O., Sauvage, J.-F., Mugnier, L.~M., Dohlen, K., \& Vigan,
  A. 2019, Mon.\ Not.\ R.\ Astron.\ Soc., 488, 4307

\bibitem[{{Kasper} {et~al.}(2008){Kasper}, {Beuzit}, {Verinaud}, {Yaitskova},
  {Baudoz}, {Boccaletti}, {Gratton}, {Hubin}, {Kerber}, {Roelfsema}, {Schmid},
  {Thatte}, {Dohlen}, {Feldt}, {Venema}, \& {Wolf}}]{Kasper2008}
{Kasper}, M.~E., {Beuzit}, J.-L., {Verinaud}, C., {et~al.} 2008, in \procspie,
  Vol. 7015, Adaptive Optics Systems, 70151S

\bibitem[{{Krist} {et~al.}(2010){Krist}, {Balasubramanian}, {Muller},
  {Shaklan}, {Kelly}, {Wilson}, {Beichman}, {Serabyn}, {Mao}, {Echternach},
  {Trauger}, \& {Liewer}}]{Krist2010}
{Krist}, J.~E., {Balasubramanian}, K., {Muller}, R.~E., {et~al.} 2010, in
  \procspie, Vol. 7731, Space Telescopes and Instrumentation 2010: Optical,
  Infrared, and Millimeter Wave, 77313J

\bibitem[{{Leboulleux} {et~al.}(2017){Leboulleux}, {N'Diaye}, {Mazoyer},
  {Pueyo}, {Perrin}, {Egron}, {Choquet}, {Sauvage}, {Fusco}, \&
  {Soummer}}]{Leboulleux2017}
{Leboulleux}, L., {N'Diaye}, M., {Mazoyer}, J., {et~al.} 2017, in Society of
  Photo-Optical Instrumentation Engineers (SPIE) Conference Series, Vol. 10562,
  Society of Photo-Optical Instrumentation Engineers (SPIE) Conference Series,
  105622Z

\bibitem[{{Leboulleux} {et~al.}(2016){Leboulleux}, {N'Diaye}, {Riggs}, {Egron},
  {Mazoyer}, {Pueyo}, {Choquet}, {Perrin}, {Kasdin}, {Sauvage}, {Fusco}, \&
  {Soummer}}]{Leboulleux2016}
{Leboulleux}, L., {N'Diaye}, M., {Riggs}, A.~J.~E., {et~al.} 2016, Society of
  Photo-Optical Instrumentation Engineers (SPIE) Conference Series, Vol. 9904,
  {High-contrast imager for Complex Aperture Telescopes (HiCAT). 4. Status and
  wavefront control development}, 99043C

\bibitem[{{Leboulleux} {et~al.}(2018){Leboulleux}, {Sauvage}, {Pueyo}, {Fusco},
  {Soummer}, {Mazoyer}, {Sivaramakrishnan}, {N'Diaye}, \&
  {Fauvarque}}]{Leboulleux2018}
{Leboulleux}, L., {Sauvage}, J.-F., {Pueyo}, L., {et~al.} 2018, Journal of
  Astronomical Telescopes, Instruments, and Systems, 4, 035002

\bibitem[{{Lightsey} {et~al.}(2014){Lightsey}, {Knight}, \&
  {Golnik}}]{Lightsey2014}
{Lightsey}, P.~A., {Knight}, J.~S., \& {Golnik}, G. 2014, in \procspie, Vol.
  9143, Space Telescopes and Instrumentation 2014: Optical, Infrared, and
  Millimeter Wave, 914304

\bibitem[{{Lyot}(1932)}]{Lyot1932}
{Lyot}, B. 1932, \zap, 5, 73

\bibitem[{{Macintosh} {et~al.}(2006){Macintosh}, {Troy}, {Doyon}, {Graham},
  {Baker}, {Bauman}, {Marois}, {Palmer}, {Phillion}, {Poyneer}, {Crossfield},
  {Dumont}, {Levine}, {Shao}, {Serabyn}, {Shelton}, {Vasisht}, {Wallace},
  {Lavigne}, {Valee}, {Rowlands}, {Tam}, \& {Hackett}}]{Macintosh2006}
{Macintosh}, B., {Troy}, M., {Doyon}, R., {et~al.} 2006, in \procspie, Vol.
  6272, Society of Photo-Optical Instrumentation Engineers (SPIE) Conference
  Series, 62720N

\bibitem[{{Malbet}(1996)}]{Malbet1996}
{Malbet}, F. 1996, \aaps, 115, 161

\bibitem[{{Mandell} {et~al.}(2017){Mandell}, {Groff}, {Gong}, {Rizzo}, {Lupu},
  {Zimmerman}, {Saxena}, \& {McElwain}}]{Mandell2017}
{Mandell}, A.~M., {Groff}, T.~D., {Gong}, Q., {et~al.} 2017, in Society of
  Photo-Optical Instrumentation Engineers (SPIE) Conference Series, Vol. 10400,
  Society of Photo-Optical Instrumentation Engineers (SPIE) Conference Series,
  1040009

\bibitem[{{Martinez} {et~al.}(2016){Martinez}, {Beaulieu}, {Janin-Potiron},
  {Preis}, {Gouvret}, {Dejonghe}, {Abe}, {Spang}, {Fant{\'e}{\"i}-Caujolle},
  {Martinache}, {Belzanne}, {Marcotto}, \& {Carbillet}}]{Martinez2016}
{Martinez}, P., {Beaulieu}, M., {Janin-Potiron}, P., {et~al.} 2016, in
  \procspie, Vol. 9906, Ground-based and Airborne Telescopes VI, 99062V

\bibitem[{{Martinez} {et~al.}(2018){Martinez}, {Janin-Potiron}, {Beaulieu},
  {Gouvret}, {Dejonghe}, {Spang}, {Postnikova}, {Baudoz}, {Guyon}, {Preis},
  {Abe}, {N'Diaye}, \& {Marcotto}}]{Martinez2018}
{Martinez}, P., {Janin-Potiron}, P., {Beaulieu}, M., {et~al.} 2018, in Society
  of Photo-Optical Instrumentation Engineers (SPIE) Conference Series, Vol.
  10703, \procspie, 1070357

\bibitem[{{Mazoyer} {et~al.}(2013){Mazoyer}, {Baudoz}, {Galicher}, {Mas}, \&
  {Rousset}}]{Mazoyer2013}
{Mazoyer}, J., {Baudoz}, P., {Galicher}, R., {Mas}, M., \& {Rousset}, G. 2013,
  \aap, 557, A9

\bibitem[{{Mazoyer} {et~al.}(2018{\natexlab{a}}){Mazoyer}, {Pueyo}, {N'Diaye},
  {Fogarty}, {Zimmerman}, {Leboulleux}, {St.~Laurent}, {Soummer}, {Shaklan}, \&
  {Norman}}]{Mazoyer2018a}
{Mazoyer}, J., {Pueyo}, L., {N'Diaye}, M., {et~al.} 2018{\natexlab{a}}, \aj,
  155, 7

\bibitem[{{Mazoyer} {et~al.}(2018{\natexlab{b}}){Mazoyer}, {Pueyo}, {N'Diaye},
  {Fogarty}, {Zimmerman}, {Soummer}, {Shaklan}, \& {Norman}}]{Mazoyer2018}
{Mazoyer}, J., {Pueyo}, L., {N'Diaye}, M., {et~al.} 2018{\natexlab{b}}, \aj,
  155, 8

\bibitem[{{Mennesson} {et~al.}(2016){Mennesson}, {Gaudi}, {Seager}, {Cahoy},
  {Domagal-Goldman}, {Feinberg}, {Guyon}, {Kasdin}, {Marois}, {Mawet},
  {Tamura}, {Mouillet}, {Prusti}, {Quirrenbach}, {Robinson}, {Rogers},
  {Scowen}, {Somerville}, {Stapelfeldt}, {Stern}, {Still}, {Turnbull}, {Booth},
  {Kiessling}, {Kuan}, \& {Warfield}}]{Mennesson2016}
{Mennesson}, B., {Gaudi}, S., {Seager}, S., {et~al.} 2016, in \procspie, Vol.
  9904, Space Telescopes and Instrumentation 2016: Optical, Infrared, and
  Millimeter Wave, 99040L

\bibitem[{{Moriarty} {et~al.}(2018){Moriarty}, {Brooks}, {Soummer}, {Perrin},
  {Comeau}, {Brady}, {Gontrum}, \& {Petrone}}]{Moriarty2018}
{Moriarty}, C., {Brooks}, K., {Soummer}, R., {et~al.} 2018, in

\bibitem[{Mugnier {et~al.}(2006)Mugnier, Blanc, \& Idier}]{Mugnier2006}
Mugnier, L.~M., Blanc, A., \& Idier, J. 2006, in Advances in Imaging and
  Electron Physics, ed. P.~Hawkes, Vol. 141 (Elsevier), 1--76

\bibitem[{{N'Diaye} {et~al.}(2014){N'Diaye}, {Choquet}, {Egron}, {Pueyo},
  {Leboulleux}, {Levecq}, {Perrin}, {Elliot}, {Wallace}, {Hugot}, {Marcos},
  {Ferrari}, {Long}, {Anderson}, {DiFelice}, \& {Soummer}}]{NDiaye2014a}
{N'Diaye}, M., {Choquet}, E., {Egron}, S., {et~al.} 2014, Society of
  Photo-Optical Instrumentation Engineers (SPIE) Conference Series, Vol. 9143,
  {High-contrast Imager for Complex Aperture Telescopes (HICAT): II. Design
  overview and first light results}, 914327

\bibitem[{{N'Diaye} {et~al.}(2013{\natexlab{a}}){N'Diaye}, {Choquet}, {Pueyo},
  {Elliot}, {Perrin}, {Wallace}, {Groff}, {Carlotti}, {Mawet}, {Sheckells},
  {Shaklan}, {Macintosh}, {Kasdin}, \& {Soummer}}]{NDiaye2013}
{N'Diaye}, M., {Choquet}, E., {Pueyo}, L., {et~al.} 2013{\natexlab{a}}, in
  \procspie, Vol. 8864, Techniques and Instrumentation for Detection of
  Exoplanets VI, 88641K

\bibitem[{{N'Diaye} {et~al.}(2013{\natexlab{b}}){N'Diaye}, {Dohlen}, {Fusco},
  \& {Paul}}]{NDiaye2013a}
{N'Diaye}, M., {Dohlen}, K., {Fusco}, T., \& {Paul}, B. 2013{\natexlab{b}},
  \aap, 555, A94

\bibitem[{{N'Diaye} {et~al.}(2015){N'Diaye}, {Mazoyer}, {Choquet}, {Pueyo},
  {Perrin}, {Egron}, {Leboulleux}, {Levecq}, {Carlotti}, {Long}, {Lajoie}, \&
  {Soummer}}]{NDiaye2015a}
{N'Diaye}, M., {Mazoyer}, J., {Choquet}, {\'E}., {et~al.} 2015, Society of
  Photo-Optical Instrumentation Engineers (SPIE) Conference Series, Vol. 9605,
  {High-contrast imager for complex aperture telescopes (HiCAT): 3. first lab
  results with wavefront control}, 96050I

\bibitem[{Paul(2014)}]{Paul2014}
Paul, B. 2014, Theses, {Universit{\'e} d'Aix-Marseille}

\bibitem[{{Paul} {et~al.}(2013{\natexlab{a}}){Paul}, {Mugnier}, {Sauvage},
  {Ferrari}, \& {Dohlen}}]{Paul2013a}
{Paul}, B., {Mugnier}, L.~M., {Sauvage}, J.~F., {Ferrari}, M., \& {Dohlen}, K.
  2013{\natexlab{a}}, Optics Express, 21, 31751

\bibitem[{{Paul} {et~al.}(2013{\natexlab{b}}){Paul}, {Sauvage}, \&
  {Mugnier}}]{Paul2013}
{Paul}, B., {Sauvage}, J.-F., \& {Mugnier}, L.~M. 2013{\natexlab{b}}, \aap,
  552, A48

\bibitem[{{Paul} {et~al.}(2014){Paul}, {Sauvage}, {Mugnier}, {Dohlen}, {Petit},
  {Fusco}, {Mouillet}, {Beuzit}, \& {Ferrari}}]{Paul2014a}
{Paul}, B., {Sauvage}, J.~F., {Mugnier}, L.~M., {et~al.} 2014, \aap, 572, A32

\bibitem[{{Perrin} {et~al.}(2018){Perrin}, {Pueyo}, {Van Gorkom}, {Brooks},
  {Rajan}, {Girard}, \& {Lajoie}}]{Perrin2018}
{Perrin}, M.~D., {Pueyo}, L., {Van Gorkom}, K., {et~al.} 2018, in \procspie,
  Vol. 10698, Society of Photo-Optical Instrumentation Engineers (SPIE)
  Conference Series

\bibitem[{{Pueyo} {et~al.}(2017){Pueyo}, {Zimmerman}, {Bolcar}, {Groff},
  {Stark}, {Ruane}, {Jewell}, {Soummer}, {St.~Laurent}, {Wang}, {Redding},
  {Mazoyer}, {Fogarty}, {Juanola-Parramon}, {Domagal-Goldman}, {Roberge},
  {Guyon}, \& {Mandell}}]{Pueyo2017}
{Pueyo}, L., {Zimmerman}, N., {Bolcar}, M., {et~al.} 2017, in Society of
  Photo-Optical Instrumentation Engineers (SPIE) Conference Series, Vol. 10398,
  Society of Photo-Optical Instrumentation Engineers (SPIE) Conference Series,
  103980F

\bibitem[{{Quanz} {et~al.}(2015){Quanz}, {Crossfield}, {Meyer}, {Schmalzl}, \&
  {Held}}]{Quanz2015}
{Quanz}, S.~P., {Crossfield}, I., {Meyer}, M.~R., {Schmalzl}, E., \& {Held}, J.
  2015, International Journal of Astrobiology, 14, 279

\bibitem[{{Rousset}(1999)}]{Rousset1999}
{Rousset}, G. 1999, {Wave-front sensors}, ed. F.~{Roddier}

\bibitem[{{Sauvage} {et~al.}(2012){Sauvage}, {Mugnier}, {Paul}, \&
  {Villecroze}}]{Sauvage2012a}
{Sauvage}, J.-F., {Mugnier}, L., {Paul}, B., \& {Villecroze}, R. 2012, Optics
  Letters, 37, 4808

\bibitem[{{Sivaramakrishnan} {et~al.}(2001){Sivaramakrishnan}, {Koresko},
  {Makidon}, {Berkefeld}, \& {Kuchner}}]{Sivaramakrishnan2001}
{Sivaramakrishnan}, A., {Koresko}, C.~D., {Makidon}, R.~B., {Berkefeld}, T., \&
  {Kuchner}, M.~J. 2001, \apj, 552, 397

\bibitem[{{Soummer} {et~al.}(2018){Soummer}, {Brady}, {Brooks}, {Comeau},
  {Choquet}, {Dillon}, {Egron}, {Gontrum}, {Hagopian}, {Laginja}, {Leboulleux},
  {Perrin}, {Petrone}, {Pueyo}, {Mazoyer}, {N'Diaye}, {Riggs}, {Shiri},
  {Sivaramakrishnan}, {St. Laurent}, {Valenzuela}, \&
  {Zimmerman}}]{Soummer2018}
{Soummer}, R., {Brady}, G.~R., {Brooks}, K., {et~al.} 2018, in

\bibitem[{{Soummer} {et~al.}(2019){Soummer}, {Laginja}, {Will},
  {Juanola-Parramon}, {Petrone}, {Brady}, {Noss}, {Perrin}, {Fowler}, {Kurtz},
  {St. Laurent}, {Fogarty}, {McChesney}, {Scott}, {Brooks}, {Comeau},
  {Ferrari}, {Gontrum}, {Hagopian}, {Hugot}, {Leboulleux}, {Mazoyer},
  {Mugnier}, {N'Diaye}, {Pueyo}, {Sauvage}, {Shiri}, {Sivaramakrishnan},
  {Valenzuela}, \& {Zimmerman}}]{Soummer2019}
{Soummer}, R., {Laginja}, I., {Will}, S., {et~al.} 2019, in Society of
  Photo-Optical Instrumentation Engineers (SPIE) Conference Series, Vol. 11117,
  \procspie

\bibitem[{{Spergel} {et~al.}(2013){Spergel}, {Gehrels}, {Breckinridge},
  {Donahue}, {Dressler}, {Gaudi}, {Greene}, {Guyon}, {Hirata}, {Kalirai},
  {Kasdin}, {Moos}, {Perlmutter}, {Postman}, {Rauscher}, {Rhodes}, {Wang},
  {Weinberg}, {Centrella}, {Traub}, {Baltay}, {Colbert}, {Bennett},
  {Kiessling}, {Macintosh}, {Merten}, {Mortonson}, {Penny}, {Rozo},
  {Savransky}, {Stapelfeldt}, {Zu}, {Baker}, {Cheng}, {Content}, {Dooley},
  {Foote}, {Goullioud}, {Grady}, {Jackson}, {Kruk}, {Levine}, {Melton},
  {Peddie}, {Ruffa}, \& {Shaklan}}]{Spergel2013}
{Spergel}, D., {Gehrels}, N., {Breckinridge}, J., {et~al.} 2013, ArXiv e-prints
  [\eprint[arXiv]{1305.5422}]

\bibitem[{{Stahl} {et~al.}(2015){Stahl}, {Shaklan}, \& {Stahl}}]{Stahl2015}
{Stahl}, M.~T., {Shaklan}, S.~B., \& {Stahl}, H.~P. 2015, in \procspie, Vol.
  9605, Techniques and Instrumentation for Detection of Exoplanets VII, 96050P

\bibitem[{{Stark} {et~al.}(2016){Stark}, {Cady}, {Clampin}, {Domagal-Goldman},
  {Lisman}, {Mandell}, {McElwain}, {Roberge}, {Robinson}, {Savransky},
  {Shaklan}, \& {Stapelfeldt}}]{Stark2016}
{Stark}, C.~C., {Cady}, E.~J., {Clampin}, M., {et~al.} 2016, in \procspie, Vol.
  9904, Space Telescopes and Instrumentation 2016: Optical, Infrared, and
  Millimeter Wave, 99041U

\bibitem[{{Stevenson} {et~al.}(2016){Stevenson}, {Lewis}, {Bean}, {Beichman},
  {Fraine}, {Kilpatrick}, {Krick}, {Lothringer}, {Mandell}, {Valenti}, {Agol},
  {Angerhausen}, {Barstow}, {Birkmann}, {Burrows}, {Charbonneau}, {Cowan},
  {Crouzet}, {Cubillos}, {Curry}, {Dalba}, {de Wit}, {Deming}, {D{\'e}sert},
  {Doyon}, {Dragomir}, {Ehrenreich}, {Fortney}, {Garc{\'{\i}}a Mu{\~n}oz},
  {Gibson}, {Gizis}, {Greene}, {Harrington}, {Heng}, {Kataria}, {Kempton},
  {Knutson}, {Kreidberg}, {Lafreni{\`e}re}, {Lagage}, {Line}, {Lopez-Morales},
  {Madhusudhan}, {Morley}, {Rocchetto}, {Schlawin}, {Shkolnik}, {Shporer},
  {Sing}, {Todorov}, {Tucker}, \& {Wakeford}}]{Stevenson2016}
{Stevenson}, K.~B., {Lewis}, N.~K., {Bean}, J.~L., {et~al.} 2016, \pasp, 128,
  094401

\bibitem[{{The LUVOIR Team}(2019)}]{LUVOIR2019}
{The LUVOIR Team}. 2019, arXiv e-prints, arXiv:1912.06219

\bibitem[{{Traub}(2003)}]{Traub2003b}
{Traub}, W.~A. 2003, Astronomical Society of the Pacific Conference Series,
  Vol. 291, {Extrasolar Planets and Biomarkers}, ed. K.~R. {Sembach}, J.~C.
  {Blades}, G.~D. {Illingworth}, \& J.~{Kennicutt}, Robert~C., 117

\bibitem[{{Wang} \& {Vaughan}(1988)}]{Wang1988}
{Wang}, Y. \& {Vaughan}, A.~H. 1988, \ao, 27, 27

\bibitem[{{Yaitskova} {et~al.}(2003){Yaitskova}, {Dohlen}, \&
  {Dierickx}}]{Yaitskova2003}
{Yaitskova}, N., {Dohlen}, K., \& {Dierickx}, P. 2003, Journal of the Optical
  Society of America A, 20, 1563

\end{thebibliography}

\end{document}